\documentclass[review]{elsarticle}
\usepackage{caption}
\usepackage{subcaption}
\usepackage{lineno,hyperref}
\usepackage{amsmath}
\usepackage{amssymb}
\usepackage{multirow}
\usepackage{svg}
\usepackage{siunitx}

\modulolinenumbers[5]

\journal{Journal of \LaTeX\ Templates}

\begin{document}

\begin{frontmatter}

\title{A camera system for real-time optical calibration of water-based neutrino telescopes}

\author[tdli]{Wei Tian}
%\ead{tianwei1997@sjtu.edu.cn}
\author[sjtu]{Wei Zhi}
\author[sjtu]{Qiao Xue}
\author[tdli]{Wenlian Li}
\author[sjtu]{Zhenyu Wei}
\author[sjtu]{Fan Hu}
\author[sjtu]{Qichao Chang}
\author[sjtu]{MingXin Wang}
\author[sjtu]{Zhengyang Sun}
\author[sjtu-chuan]{Xiaohui Liu}
\author[tdli]{Ziping Ye}
\author[ustc]{Peng Miao}
\author[sjtu-chuan,sjtu-Yazhou]{Xinliang Tian}
\author[tdli,sjtu]{Jianglai Liu}
\author[tdli,sjtu]{Donglian Xu\corref{corresponding_author}}
\ead{donglianxu@sjtu.edu.cn}
\cortext[corresponding_author]{Corresponding author}

\address[tdli]{Tsung-Dao Lee Institute, Shanghai Jiao Tong University, Shanghai 201210, China}
\address[sjtu]{School of Physics and Astronomy, Shanghai Jiao Tong University, Key Laboratory for Particle Astrophysics and Cosmology (MoE), Shanghai Key Laboratory for Particle Physics and Cosmology, Shanghai 200240, China} 

\address[pku]{Department of Astronomy, School of Physics, Peking University, Beijing 100871, China}
\address[ustc]{Department of Modern Physics, University of Science and Technology of China, Hefei 230026, China}
\address[sjtu-chuan]{State Key Laboratory of Ocean Engineering, School of Naval Architecture Ocean and Civil Engineering, Shanghai Jiao Tong University, Shanghai, 200240, China}
\address[sjtu-Yazhou]{Shanghai Jiao Tong University Sanya Yazhou Bay Institute of Deep Sea Technology, Sanya, 572024, China}

\begin{abstract}
Calibrating the optical properties within the detection medium of a neutrino telescope is crucial for determining its angular resolution and energy scale. For the next generation of neutrino telescopes planned to be constructed in deep water, such as the TRopIcal DEep-sea Neutrino Telescope (TRIDENT), there are additional challenges due to the dynamic nature and potential non-uniformity of the water medium. This necessitates a real-time optical calibration system distributed throughout the large detector array. This study introduces a custom-designed CMOS camera system equipped with rapid image processing algorithms, providing a real-time optical calibration method for TRIDENT and other similar projects worldwide. In September 2021, the TRIDENT Pathfinder experiment (TRIDENT Explorer, T-REX for short) successfully deployed this camera system in the West Pacific Ocean at a depth of 3420 meters. Within 30 minutes, about 3000 images of the T-REX light source were captured, allowing for the in-situ measurement of seawater attenuation and absorption lengths under three wavelengths. This deep-sea experiment for the first time showcased a technical demonstration of a functioning camera calibration system in a dynamic neutrino telescope site, solidifying a substantial part of the calibration strategies for the future TRIDENT project. 

\end{abstract}


\end{frontmatter}

\section{Introduction}
\label{sec:intro}
%\subsection{Brief Introduction of TRIDENT}

High-energy neutrinos have ushered in a new era of astrophysics, serving as unique messengers to unveil the mysteries surrounding the origins of cosmic rays. Following the exciting discovery and subsequent studies of astrophysical neutrinos by IceCube \cite{IceCube, IceCube_TXS_flares:2018, IceCube_NGC1068:2022, IceCube_Milky_way}, TRIDENT is proposed as a next-generation neutrino telescope and planned to be constructed at a depth of approximately 3500 meters in the Western Pacific Ocean, north of the South China Sea \cite{TRIDENT}. TRIDENT aims to boost the search for high-energy neutrino sources and optimize for all-flavor neutrino detection, paving the way for further investigations into astrophysical neutrino production mechanisms within their sources and the realm of new physics \cite{neutrino_flavor, Flavor_Physics, quantum_gravity, Flavor_dark_matter}. The equatorial location of TRIDENT also allows TRIDENT to provide complementary neutrino sky coverage alongside IceCube. In September 2021, the TRIDENT Pathfinder experiment was conducted as the first expedition for site selection. It successfully collected data on oceanographic conditions and natural seawater radioactivity, with a particular emphasis on measuring the optical properties of deep-sea water at a depth of 3420 meters. Additionally, the expedition tested ocean engineering technology through the deployment of a detector prototype.

A neutrino telescope detects high-energy neutrinos by observing the Cherenkov photons emitted by the relativistic secondary charged particles generated from the neutrino interaction vertices. The direction and energy information of the primary neutrinos can be reconstructed by analyzing the quantity and arrival times of the Cherenkov photons received by the detector array. These photons, however, can undergo random absorption and scattering processes in the medium before being detected \cite{mobley1994light}. Absorption processes can cause the loss of photons as their energy is converted into undetectable atomic heat, while scattering results in photon deflection, leading to blurred arrival times recorded by the detector.

Therefore, accurate calibration of the optical properties of the medium is crucial for optimizing the performance of a neutrino telescope, especially for its angular resolution and energy scale. In the case of water-based neutrino telescopes like TRIDENT, pure deep-sea water is expected to have less photon scattering effect compared to glacial ice, but it also presents additional technical challenges. These challenges include the dynamic nature of water, which can cause time variations in optical parameters, and the large volume of water within the detector array, potentially leading to local nonuniformities. To address these effects and ensure accurate measurements, the development of a fast real-time calibration system distributed among the detector array is necessary.

%Therefore, the future design and final performance of a neutrino telescope, including its energy threshold and angular resolution, significantly correlate with the in-situ optical properties. 

Various optical calibration techniques have been employed in existing neutrino telescopes, such as IceCube and ANTARES \cite{ANTARES}, and others under development, including KM3NeT \cite{KM3NeT}, Baikal-GVD \cite{Baikal_GVD}, and P-ONE \cite{P-ONE}. Most of these experiments use Photomultiplier Tubes (PMTs) coupled with pulsing light sources for their calibration strategies. Optical parameters are measured by analyzing the received photon amounts at different distances or fitting the accumulated photon arrival time distribution from the PMT data \cite{ANTARES2004, straw-a, LAMS}. This approach is sensitive to photon absorption and scattering effects due to the single-photon detection capability and precise time measurement of PMTs. However, it also requires relatively long-duration data accumulation under the single-photon detection mode to obtain sufficient photon statistics, which relies on precise time synchronization between PMTs and the pulsing light source. Moreover, the dynamic underwater environment can introduce additional background noise to the PMT signals \cite{TRIDENT_PMT}, posing extra challenges for single-photon selection. Additionally, transient deep-sea currents or bioactivity could lead to short-term variations in optical properties, potentially rising difficulties for the long-duration photon accumulation process.

%In conclusion, this strategy usually requires high robustness of the detection systems, and its measurement accuracy can be constrained by the statistics and time response characteristics of both the PMTs and the pulsing light source.

Some other alternative calibration techniques have also been explored, such as specialized laser apparatus like AC9 \cite{AC9} and Baikal-5D \cite{Baikal_5D}. These instruments achieve high accuracy through mature calibration strategies and well-calibrated laser emitters and receivers. However, they are typically used for localized optical measurements and require independent power and data transmission systems, posing challenges for real-time calibration across different regions within large detector arrays.

CMOS cameras, a well-established technology, offer great potential for optical calibration in neutrino telescopes. Their compact size and low power consumption facilitate easy integration into optical detection modules, enabling calibration coverage across entire detector arrays. Moreover, cameras can achieve rapid calibration by operating with a stable light source, allowing for the accumulation of sufficient photon statistics even with milliseconds of exposure. Ongoing related research is actively exploring the use of CMOS cameras in neutrino telescopes for tasks such as bioluminescence monitoring and specific optical calibration of glacial ice \cite{ANTARES_bio, KM3NeT_camera, straw_b, PONE-phd_thesis, IceCube_gen2_camera, IceCube-gen2_ICRC2023}.

In this paper, we present a real-time calibration system based on CMOS cameras, designed for fast, real-time measurement of the optical properties of deep-sea water. 
The system has been successfully demonstrated by the T-REX mission for its robustness and accuracy \cite{TRIDENT}. The paper is structured as follows: Section \ref{Taking photos} introduces the experimental settings and hardware design of the T-REX and the camera system. Section \ref{models} discusses the photon propagation models from previous experiments and introduces a refined optical model for the T-REX. In Section \ref{Analysis methods}, we present two methods for analyzing data from the camera system to measure attenuation and absorption lengths. Section \ref{sec:Calibration} presents the calibration strategies for the camera system. Finally, Section \ref{conclusion_outlook} provides a summary and outlook.
%In this case, a real-time and widely-covered calibration system is needed for a deap-water neutrino telescope like TRIDENT to deal with the fluidity and nonuniformity of the large volume of water medium. lattice-distributed

%, some pilot experiments have been operated for the existing or under-development neutrino telescopes in glacial ice (IceCube, 1998\cite{}) or deep water(ANTARES,2004\cite{} KM3NeT\cite{}, Baikal-GVD\cite{} and P-ONE\cite{}).  Most of these experiments used Photon Multiplier Tube (PMT) to receive the photons from pulsing LED and obtained the optical parameters by fitting the photon numbers and their arrival times. \cite{}  依赖于光子的到达时间的这种测量方式，需要比较长的测量时间才能获得足够统计量的单光子信号，并且对于单光子信号的产生与接收时间的准确性有很高的要求。like IceCube\cite{} in glacial ice and other water-based ones such as KM3NeT\cite{}, Baikal-GVD\cite{} and P-ONE\cite{},

%这里要提到Ice-cube gen2

\section{Design of the T-REX and the camera system}
\label{Taking photos}%xq要引用这个章所以加了这个label，请不要删

\begin{figure}[!th]
    \centering
    \includegraphics[width=0.7\linewidth]{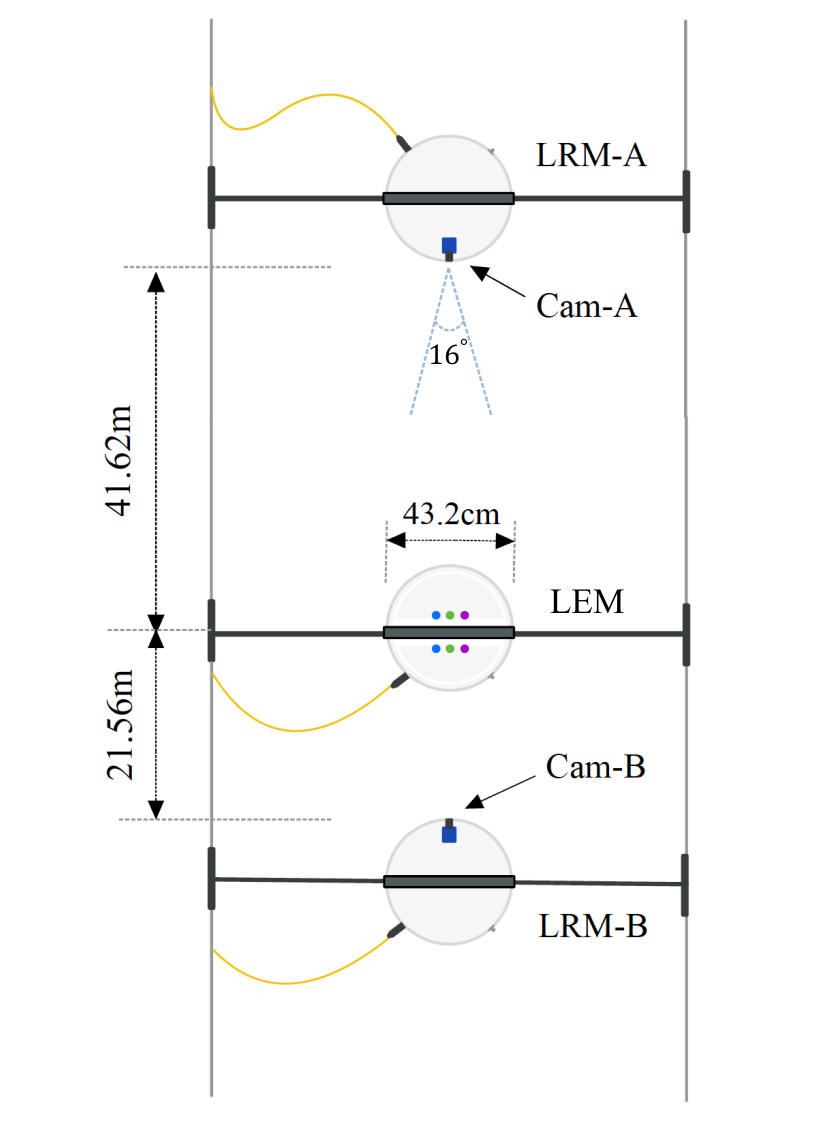}
    \caption{A schematic illustration of the T-REX.} 
    \label{Figure: TRIDENT Pathfinder experiment apparatus}
\end{figure} 

The T-REX apparatus comprises a light emission module (LEM) and two light receiver modules (LRMs): LRM-A and LRM-B, as shown in Figure \ref{Figure: TRIDENT Pathfinder experiment apparatus}. LRM-A is positioned at a vertical distance of $21.56\pm0.02~\mathrm{m}$ from the LEM, while LRM-B at $41.62\pm 0.04~\mathrm{m}$. Each LRM is equipped with a CMOS camera (Cam-A or Cam-B) and three 3-inch PMTs, capable of detecting light signals from the central LEM at two different distances. The LEM operates in two emission modes: steady mode and pulsing mode. The steady mode ensures consistent illumination for capturing clear images of the LEM by the cameras, while the pulsing mode is adapted for the PMT system \cite{TRIDENT_PMT}. To cover the detectable waveband of Cherenkov radiation in water, the steady mode employs LEDs with wavelengths of 405 nm, 460 nm, and 525 nm.

Throughout the T-REX deployment process, the entire apparatus was hoisted by the research vessel using an umbilical cable, maintaining straight due to self-gravity in the water. During the descent at water depths of 1200 m and 2021 m, the camera system conducted periodic tests for image capture. Upon reaching the target depth of 3420 m, the camera system was activated and carried out an image acquisition process lasting approximately 30 minutes. It successfully captured about 3000 images of the LEM with all three wavelengths. Throughout the retrieval procedure, both the camera system and environmental monitoring sensors inside the LRMs remained operational, collecting data at various water depths.
% The whole image taking process lasts about 30 minutes before the T-REX system was retrieved. During the retrieval, the camera system and environment monitoring sensors still stays in a working state and record corresponding data. 
\begin{figure}[!ht]
\centering % \begin{center}/\end{center} takes some additional vertical space
\includegraphics[width=.8\textwidth]{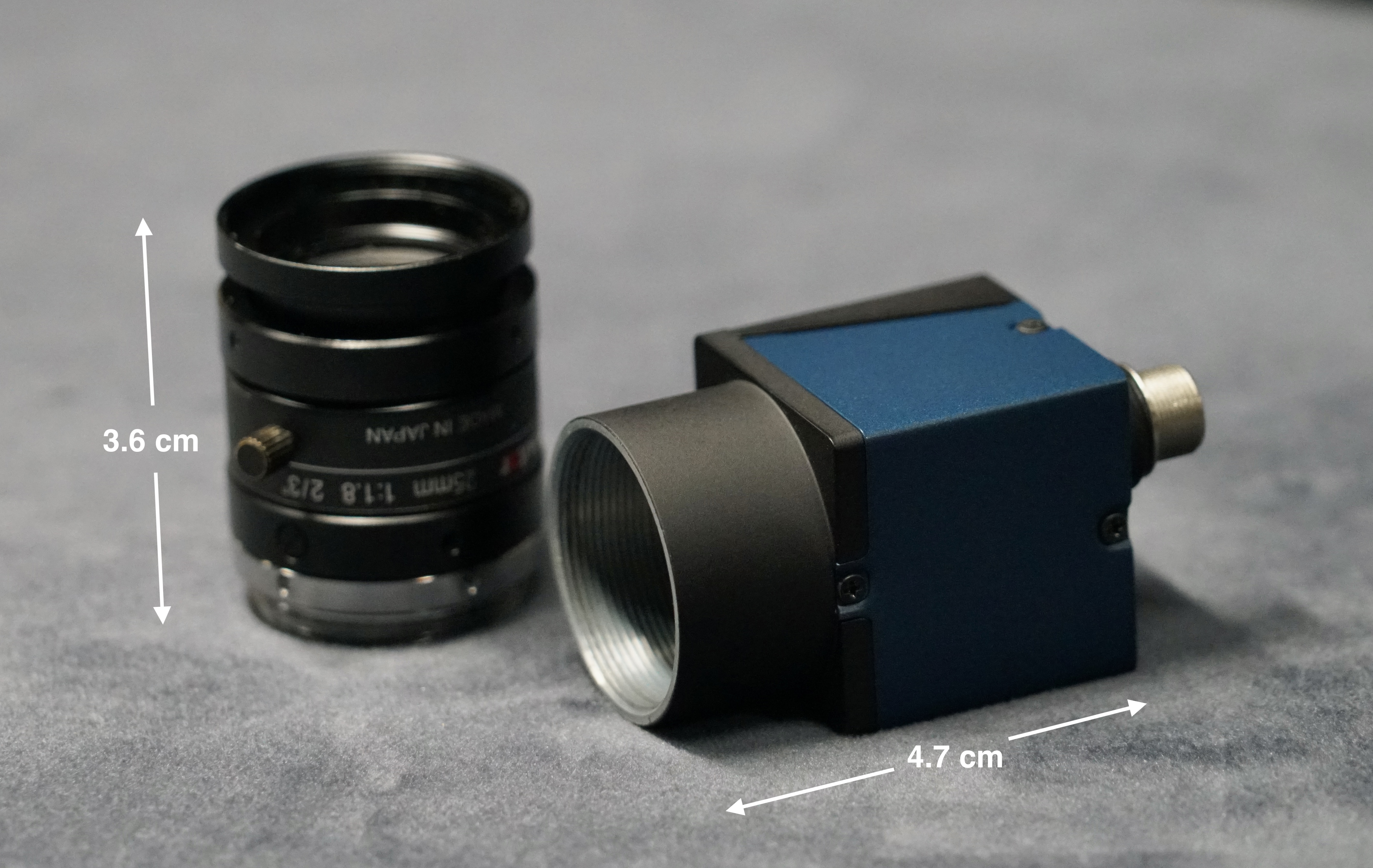}
% "\includegraphics" from the "graphicx" permits to crop (trim+clip)
% and rotate (angle) and image (and much more)
\caption{\label{camera_lens} An illustration of the camera system's size.}
\end{figure}

\begin{figure}[!ht]
\centering % \begin{center}/\end{center} takes some additional vertical space
\includegraphics[width=.8\textwidth]{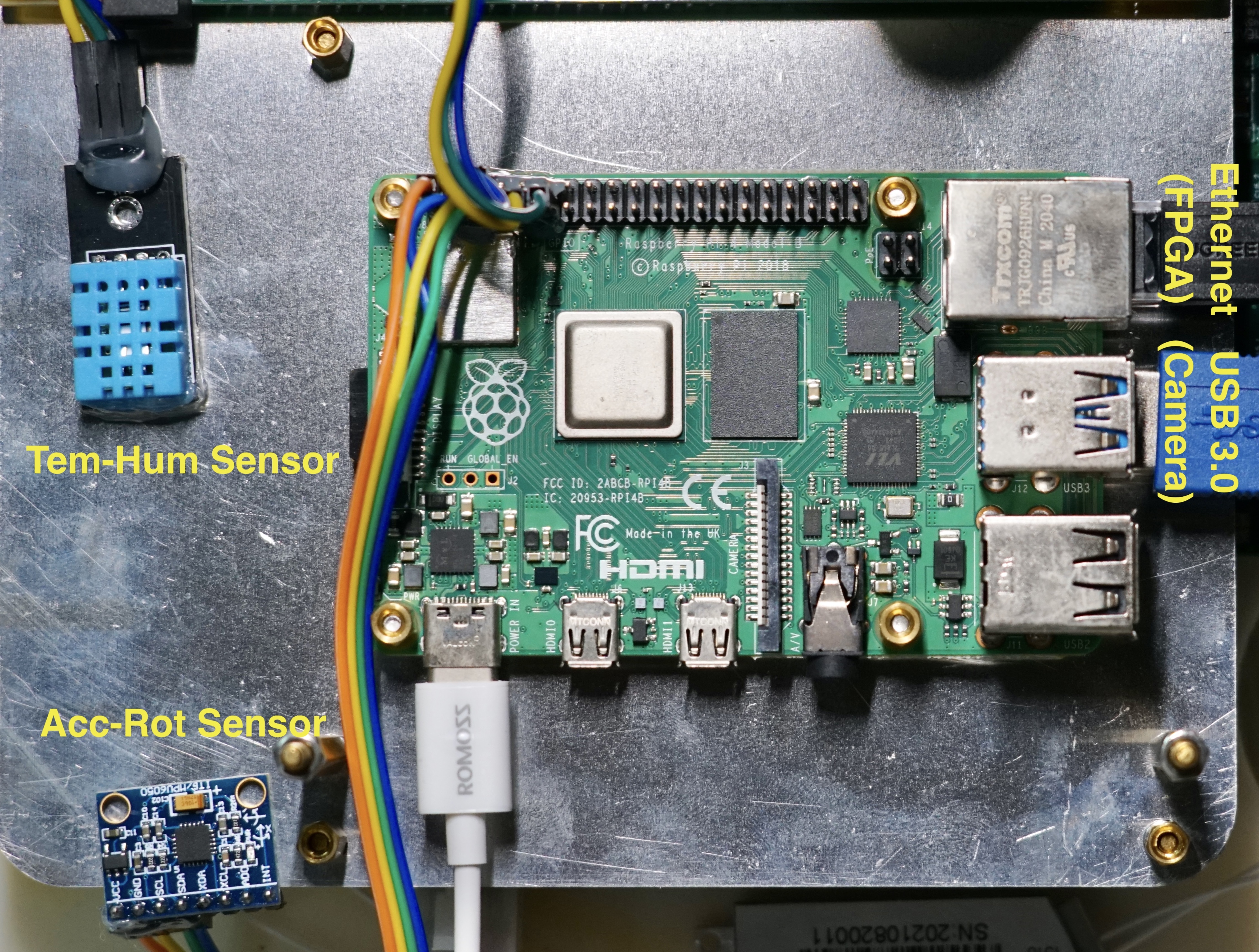}
% "\includegraphics" from the "graphicx" permits to crop (trim+clip)
% and rotate (angle) and image (and much more)
\caption{\label{hardware2} The control board of the camera system and assistant sensors.}
\end{figure}

%\section{Hardware of Camera System in T-REX}

The camera system, as depicted in Figure \ref{camera_lens}, consists of a monochromatic 5 million-pixel camera paired with a 25 mm-focal-length lens, providing a viewing angle of approximately $16^\circ$ in water. Its automated control is facilitated by a Raspberry Pi module, which supplies a 5.0 V driving voltage and enables data transmission through its USB 3.0 port. Additionally, the Raspberry Pi module is equipped with temperature-humidity and acceleration-rotation sensors. These sensors monitor the internal environment and dynamic state of the LRM, providing supplementary data such as the leaning angle and the rotation speed along three axes. Upon activation of the Raspberry Pi module, both the camera and the sensors automatically start data acquisition. The captured images and sensor readings are then transmitted in real-time back to the research vessel from the deep sea via the optical fiber within the umbilical cable. Furthermore, a duplicate copy of the data is stored on a local Secure Digital Memory Card (SD card) as a safeguard against potential transmission loss.

To accommodate uncertain underwater conditions and variations in the intrinsic light intensities of the LEM at three wavelengths, the camera system was configured with a wide range of exposure time and gain combinations during the experiment. The exposure time settings included 0.02 s, 0.05 s, 0.07 s, 0.11 s, 0.2 s, 0.5 s, and 1.0 s, while the gain settings ranged from 2 dB to 20 dB with an increment of 2 dB. This configuration ensured that the captured image gray values remained within the linear response range of the camera, avoiding potential saturation. For each exposure time and gain combination, the camera recorded 20 images as repeated measurements.

The lower exposure time limit, 0.02 s, was set as a precaution against potential sway or disturbances of the LRM caused by underwater currents. It aimed to maintain image motion blur below 0.7 pixels, considering a LEM moving speed of 0.1 m/s within the field of view and a measurement distance of 20 m. In practice, measurements of current speed \cite{TRIDENT} at 3420 m indicated speeds of less than 10 cm/s, and data from dynamic sensors suggested minimal current disturbances during the deployment process, as shown in \ref{App A}. Exposure times of 0.05 s, 0.07 s, and 0.11 s were determined based on LEM calibration results for three wavelengths \cite{TRIDENT_light_source}. Furthermore, exposure times of 0.2 s, 0.5 s, and 1.0 s primarily served for environmental monitoring, recording potential biological activity, and capturing deployment dynamics.

\section{Photon propagation models in less-scattering water}
\label{models}
% In the above-mentioned pilot experiments to decode the water optical property 

In this section, we present the adapted photon propagation models in deep-sea water based on the T-REX design. Considering that the optical models vary across different types of mediums and light source setups, we first give a comprehensive review for the optical models commonly used in previous similar experiments and compare their definitions of the optical parameters \cite{ANTARES2004, straw-a, LAMS, AC9, IceCube-diffusion, Baikal-1999}.

Furthermore, we aim to construct a refined model, particularly for the case of a spherical isotropic light source in a less-scattering water medium, using only canonical optical parameters. We can then incorporate experimentally determined optical parameters into future neutrino telescope simulations, 
mitigating systematic uncertainties in TRIDENT arising from optical simulations.

%如何引用一下icecube的文章，致敬一下diffusion模型

%因为之前的一些实验测量对于某些光学参数的定义都有所不同，所以这里讲会介绍
\subsection{Attenuation effect in the light beam scenario}

As a fundamental optical parameter, the attenuation length characterizes the exponential decay of the intensity of a monochromatic light beam as it traverses through a medium \cite{beer_lambert_law}: 

\begin{equation}
I(L)=I_{0} \cdot e^{-\frac{L}{_{\lambda _{\mathrm{att}} } } } 
\label{def of att}
\end{equation}

\begin{figure}[htbp]
\centering % \begin{center}/\end{center} takes some additional vertical space
\includegraphics[width=.6\textwidth]{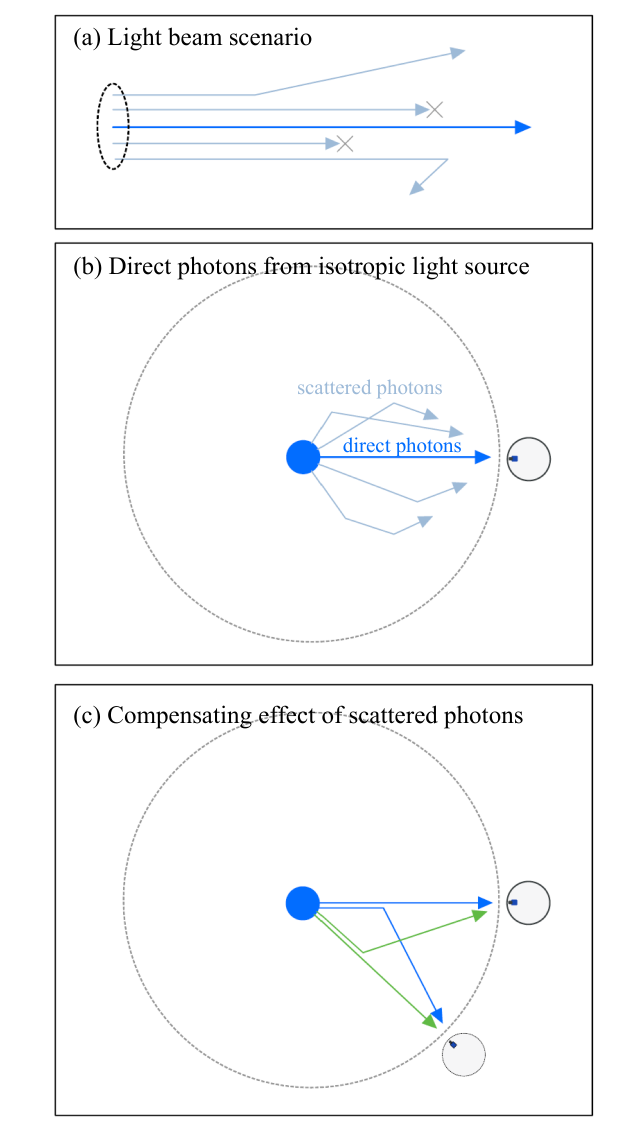}
% "\includegraphics" from the "graphicx" permits to crop (trim+clip)
% and rotate (angle) and image (and much more)
\caption{\label{optical_model} 
Illustrations of the optical models for the light beam and isotropic light source scenarios are shown. Panel (a) depicts the absorption and scattering processes in the light beam scenario, where both lead to photon loss along the initial beam direction. Panels (b) and (c) present the optical model in an isotropic pulsing light source scenario. In panel (b), all the scattered photons follow a spherical symmetrical distribution illustrated by the outer dashed line. The scattered photons, depicted by shallow blue arrows, arrive later than the direct photons, depicted by deep blue arrows, due to deflection. Panel (c) illustrates the compensating effect of scattered photons due to geometric symmetry, where the LRM placed anywhere on the sphere receives the same scattering pattern. The green and blue arrows show an example of the scattered photons compensating each other, indicating that the scattering effect does not directly lead to photon loss in this scenario.}
\end{figure}

\noindent Here, $I_\mathrm{0}$ represents the initial radiance of the light beam, and after traveling a distance $L$, the remaining radiance along the incident direction is denoted as $I(L)$. The canonical attenuation length, $\lambda_{\mathrm{att}}$, is defined as the mean free path in the medium, accounting for photon loss due to both absorption and scattering processes, as illustrated in Figure \ref{optical_model}. $\lambda_{\mathrm{att}}$ can be further decomposed into two independent components:

\begin{equation}
\frac{1}{_{\lambda _\mathrm{{att}} }} = \frac{1}{_{\lambda _\mathrm{{abs}} }} + \frac{1}{_{\lambda _\mathrm{{sca}} }}
\label{compose of att}
\end{equation}

\noindent In this equation, $\lambda_{\mathrm{abs}}$ and $\lambda_{\mathrm{sca}}$ represent the absorption length and scattering length, respectively. The scattering length encompasses the combined effects of Rayleigh and Mie scattering, which are the dominant elastic scattering processes in natural water. The physical models of Rayleigh and Mie scattering depend on the size of the scatterer, characterized by different phase functions that describe the angular distribution of scattered photons \cite{phase_function}.

%这里需要强调之所以可以独立的拆分为两项，成立的原因是因为窄束光的散射偏离和吸收都会最终的光损耗。

%introduction of eff-att / eff-abs
%In some other experiments, another optical parameter, effective attenuation length
\subsection{Absorption-scattering combination effect in the isotropic light source scenario}

For a spherical or point-like isotropic light source emitting photons in all directions, however, Formula \ref{def of att} should be modified to account for the additional geometric symmetry. 

Assuming the light source emits $N_\mathrm{0}$ photons in a short pulse, and a photon detector is placed at a distance $L$ with a small detection area $dS$. The detector will first receive those directly-arrived photons that have not been absorbed or scattered. At this moment, the scattered photons are still propagating due to deflection, so they need longer light paths to reach the detector, as illustrated in Figure \ref{optical_model}. Therefore, if we only consider the number of the directly-arrived photons, denoted as $N_{\mathrm{dir}}(L)$, it will still follow a similar exponential decrease but with an additional inverse square term due to the spherical symmetry:

\begin{equation}
N_\mathrm{{dir}}(L)=N_{0} \cdot \frac{dS}{4\pi L^2} \cdot e^{-\frac{L}{_{\lambda _{\mathrm{att}} } } } 
\label{def of att for sphe}
\end{equation}

This formula presents an experimental strategy to measure the canonical attenuation length by screening the arrival time of each photon with PMTs. In practice, separating the directly-arrived photons is challenging due to intrinsic time response uncertainties in both PMTs and the pulsing light source. Even with nanosecond-level time response uncertainty, a non-negligible portion of scattered photons, which have small scattering angles or are scattered near the detector, can mix with the directly-arrived photons. Therefore, selecting only photons with shorter arrival times from PMT data can introduce additional measurement bias. This bias is investigated in our simulation study conducted for the PMT system of the T-REX \cite{Sim_paper, TRIDENT_PMT, TRIDENT_light_source}. It also inspires the consideration of selecting directly-arrived photons not solely based on the time information but also from the angular distribution of the light field captured by the cameras. This will be fully discussed later in the section of analysis methods \ref{Analysis methods}.

For those scattered photons, they do not get filtered out like in the light beam scenario but still follow a spherical symmetry distribution. This is because even if some photons initially pointed towards the detector are deflected away, there are also some scattered photons compensating for this direction, initially emitted in another direction. Under an isotropic light source and medium assumption, these two effects balance out due to geometric symmetry. Therefore, when calculating the total number of photons, including both scattered and directly-arrived photons at distance $L$, Equation \ref{def of att for sphe} is no longer valid. Moreover, scattered photons are more likely to be absorbed due to their longer light paths compared to non-scattered ones. This absorption-scattering combination effect can lead to additional photon loss, which becomes significant when $L \gtrsim \lambda _\mathrm{{sca}}$.

To quantify the total number of detectable photons $N(L)$ at distance $L$, an effective model is commonly applied in previous experiments \cite{ANTARES2004, LAMS}:

\begin{equation}
N(L)=N_{0} \cdot \frac{dS}{4\pi L^2} \cdot e^{-\frac{L}{_{\lambda _{\mathrm{eff,att}} } } } 
\label{def of eff-att for sphe}
\end{equation}

\noindent Here, $\lambda_{\mathrm{eff,att}}$ represents the effective attenuation length, which differs from the canonical attenuation length $\lambda_{\mathrm{att}}$ in Equation \ref{def of att for sphe}. It's worth noting that $\lambda_{\mathrm{eff,att}}$ is not strictly an intrinsic parameter of the medium but rather a function dependent on $L$, $\lambda_{\mathrm{abs}}$, and $\lambda_{\mathrm{sca}}$. This dependency arises because the proportion of multi-scattered photons increases with $L$, causing the additional photon loss to vary with distance due to the absorption-scattering combination effect.
For a short measurement distance $L\ll \lambda_{\mathrm{sca}}$, the scattering effect in the medium is weak, allowing $\lambda_{\mathrm{eff,att}}$ to converge to $\lambda_{\mathrm{abs}}$.
% This approximation holds valid under the assumption of a short measurement distance and weak scattering effect in the medium, where $L\ll \lambda_{\mathrm{sca}}$, allowing $\lambda_{\mathrm{eff,att}}$ to converge to $\lambda_{\mathrm{abs}}$.
However, in water-based neutrino telescopes, the typical inter-string distance is on the same order as $\lambda_{\mathrm{sca}}$, leading to non-negligible scattering effects even though absorption typically dominates.

\subsection{A refined model for isotropic light source in less-scattering water}

To better describe the scattering effect in the medium, the effective scattering length $\lambda_{\mathrm{eff,sca}}$ has been introduced in previous works \cite{IceCube-diffusion, Baikal-1999} to represent the average effect of multiple scattering in a diffusive medium like glacial ice. Its definition combines the canonical scattering length with the typical mean scattering angle $\left \langle \mathrm{cos}\theta \right \rangle$, based on an assumption of infinite scattering times:

\begin{equation}
\lambda_\mathrm{{eff,sca}} = \frac{\lambda_\mathrm{{sca}}}{1-\left \langle \mathrm{cos}\theta \right \rangle}
\label{eff_sca}
\end{equation}

However, the deep-sea water typically has a much longer scattering length than glacial ice, so it cannot be treated as a diffusive medium with infinite scattering times among the detector array. Therefore, it's motivated to establish a model for a less-scattering water medium that describes $N(L)$ at a distance $L$, considering both the longer light paths of scattered photons and the absorption-scattering combination effect.

\begin{equation}
N(L)=N_{0} \cdot \frac{dS}{4\pi L^2} \cdot e^{-\frac{\overline{L}}{_{\lambda _{\mathrm{abs}} } } } 
\label{TW model}
\end{equation}

In this model, we use $\overline{L}$ to replace $L$, representing the weighted mean light path of all detected photons after scattering. $\overline{L}$ is an analytic function depending on $L$, the Rayleigh and Mie scattering lengths, and their phase functions. With the definition of $\overline{L}$, the absorption effect can be separated from scattering, allowing the canonical absorption length to be directly applied in the exponential decay term to account for the total photon loss. 

In \ref{App B}, we provide a detailed analytical calculation of the first and second order of $\overline{L}$. This model exhibits good numerical accuracy when fitting data from Geant4 simulations \cite{Sim_paper}. The results also demonstrate the model's validity even when $L$ is in the same order as $\lambda_{\mathrm{sca}}$. 

This refined model presents a more accurate description of the photon distribution in less-scattering deep-sea water adapted to the T-REX. As a direct application, it provides an alternative method for measuring $\lambda_{\mathrm{abs}}$ using the PMT system \cite{Sim_paper}. By counting the scattered photons within a specific arrival time bin $t_i$, corresponding to a fixed photon light path $L_i$ in water, $\lambda_{\mathrm{abs}}$ can be calculated from the ratio of the number of these photons detected by LRM-A and LRM-B.

\section{Analysis methods for optical measurements with the camera system}
\label{Analysis methods}
Based on the above study of optical models, this section presents two analysis methods established for the camera system to measure the canonical optical parameters from images captured by the LRMs: the $I_{\mathrm{center}}$ method for measuring $\lambda_{\mathrm{att}}$ and the statistical $\chi^2$ method for measuring $\lambda_{\mathrm{abs}}$ and $\lambda_{\mathrm{sca}}$.
% These images record both direct and scattered photons from the steady LEM.

\subsection{$I_{\mathrm{center}}$ method for $\lambda_{\mathrm{att}}$ measurement}
\label{I_center Method}

% Based on the previously introduced optical models, this section introduces two analysis methods to obtain optical parameters from images captured by the LRMs' cameras. These images record both direct and scattered photons from the steady LEM.

% Firstly, we present the $I_{\mathrm{center}}$ method for measuring $\lambda_{\mathrm{att}}$.
In T-REX, the cameras inside the two LRMs can be considered as pinhole cameras because of their long distances from the LEM. Therefore, each pixel of the camera records the radiance from a specific direction from the LEM. The gray value of each pixel is proportional to the number of received photons. Since the size of the LEM only occupies a $0.6^\circ$ and $1.1^ \circ$ viewing angle for Cam-A and Cam-B, respectively, the directly-arrived photons from the LEM are highly collimated and concentrated in the center pixels on the image. Conversely, the scattered photons form a scattering halo around the center pixels, with a widely-spread angular distribution. Therefore, by analyzing the gray values of the LEM centroid pixel on the images, scattered photons can be effectively excluded. Consequently, according to Formula \ref{def of att}:

\begin{subequations}
\begin{equation}
I_{A} = I_{0} \cdot f \cdot e^{-L_{A}/\lambda_{\mathrm{att}}}\\
\end{equation}
\begin{equation}
I_{B} = I_{0}' \cdot f \cdot e^{-L_{B}/\lambda_{\mathrm{att}}}\\
\label{I_center_AB}
\end{equation}
\begin{equation}
\lambda_{\mathrm{att}} = -(L_{\mathrm{A}}-L_\mathrm{{B}}) /\ln(-\frac{I_{\mathrm{A}}}{I_\mathrm{{B}}}\cdot \frac{I_{0}'}{I_{0}})
\label{I_center}
\end{equation}
\end{subequations}

\begin{figure}[!ht]
\centering % \begin{center}/\end{center} takes some additional vertical space
\includegraphics[width=.99\textwidth]{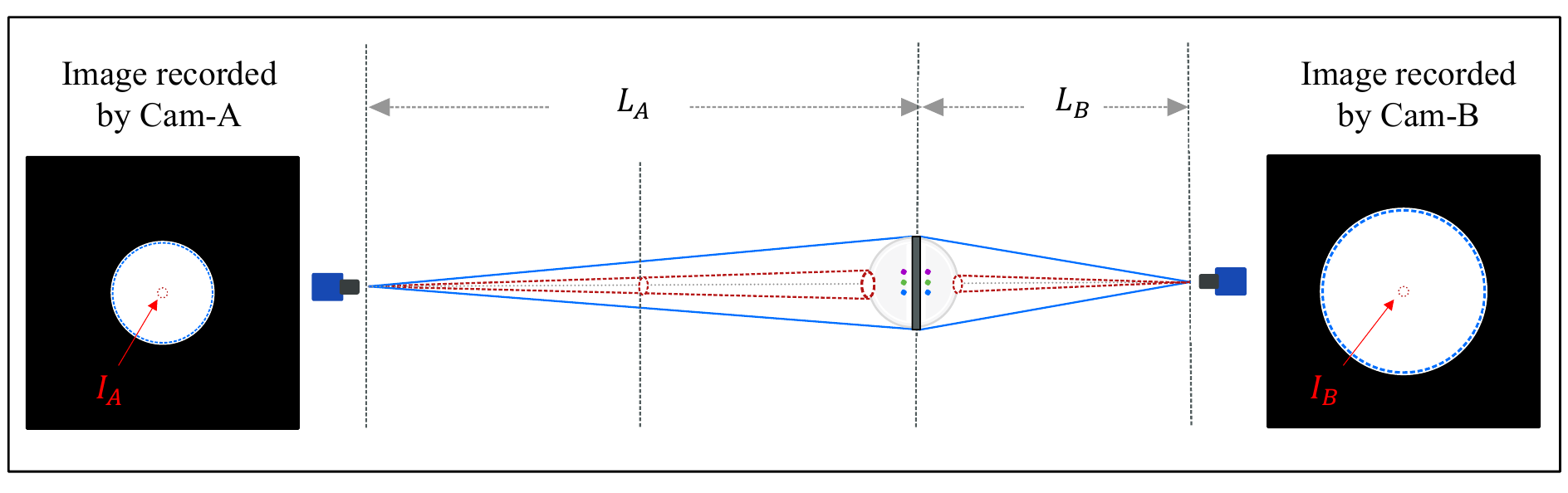}
% "\includegraphics" from the "graphicx" permits to crop (trim+clip)
% and rotate (angle) and image (and much more)
\caption{\label{fig I_center} An illustration of $I_{\mathrm{center}}$ method. The blue lines show the viewing angles occupied by the LEM for Cam-A and Cam-B, which are $0.6^\circ$ and $1.1^ \circ$. The red dashed lines show the unit solid angle of a single pixel. We analyze the mean gray values of the center pixels marked on the images to calculate $\lambda_{\mathrm{att}}$.}
\end{figure}

Here, $I_\mathrm{{A}}$ and $I_\mathrm{{B}}$ represent the mean gray values of the pixels around LEM centroid on the images recorded by Cam-A and Cam-B, respectively, with the same exposure time and gain. $L_\mathrm{{A}}$ and $L_\mathrm{{B}}$ are the measurement distances between the LEM and the cameras, and $L_{A}$ is approximately twice as long as $L_{B}$, as illustrated in Figure \ref{fig I_center}. The factor $I_{0}/I_{0}'$ denotes the ratio of light emission intensity between the two sides of the LEM, accounting for the intrinsic non-uniformity of the light source. This ratio is pre-calibrated in the laboratory before the deep-sea experiment \cite{TRIDENT_light_source}. The transmission rate $f$ accommodates for the additional photon loss at the water-glass-air interfaces, which can be canceled out in a relative measurement.

As mentioned above, the $I_{\mathrm{center}}$ method relies on the angular distribution of the light field to select direct photons, necessitating a sufficiently small viewing angle occupied by the LEM to exclude most scattered photons. Compared to screening their arrival times by PMTs, the residual scattered photons within such a small viewing angle typically amount to less than 10\%, as indicated by simulation studies \cite{Sim_paper}. Therefore, the potential measurement bias of the $I_{\mathrm{center}}$ method can be calculated to be less than 4\%, even under simulation settings where the scattering length is less than 30 m \cite{Sim_paper}.

% written as: As a result, the attenuation length can be measured according to \ref{def of att} 

% by comparing , $I_\mathrm{{A}}$ and $I_\mathrm{{B}}$, recorded by Cam-A and Cam-B, respectively:

\begin{figure}[!ht]
\centering % \begin{center}/\end{center} takes some additional vertical space
\includegraphics[width=.99\textwidth]{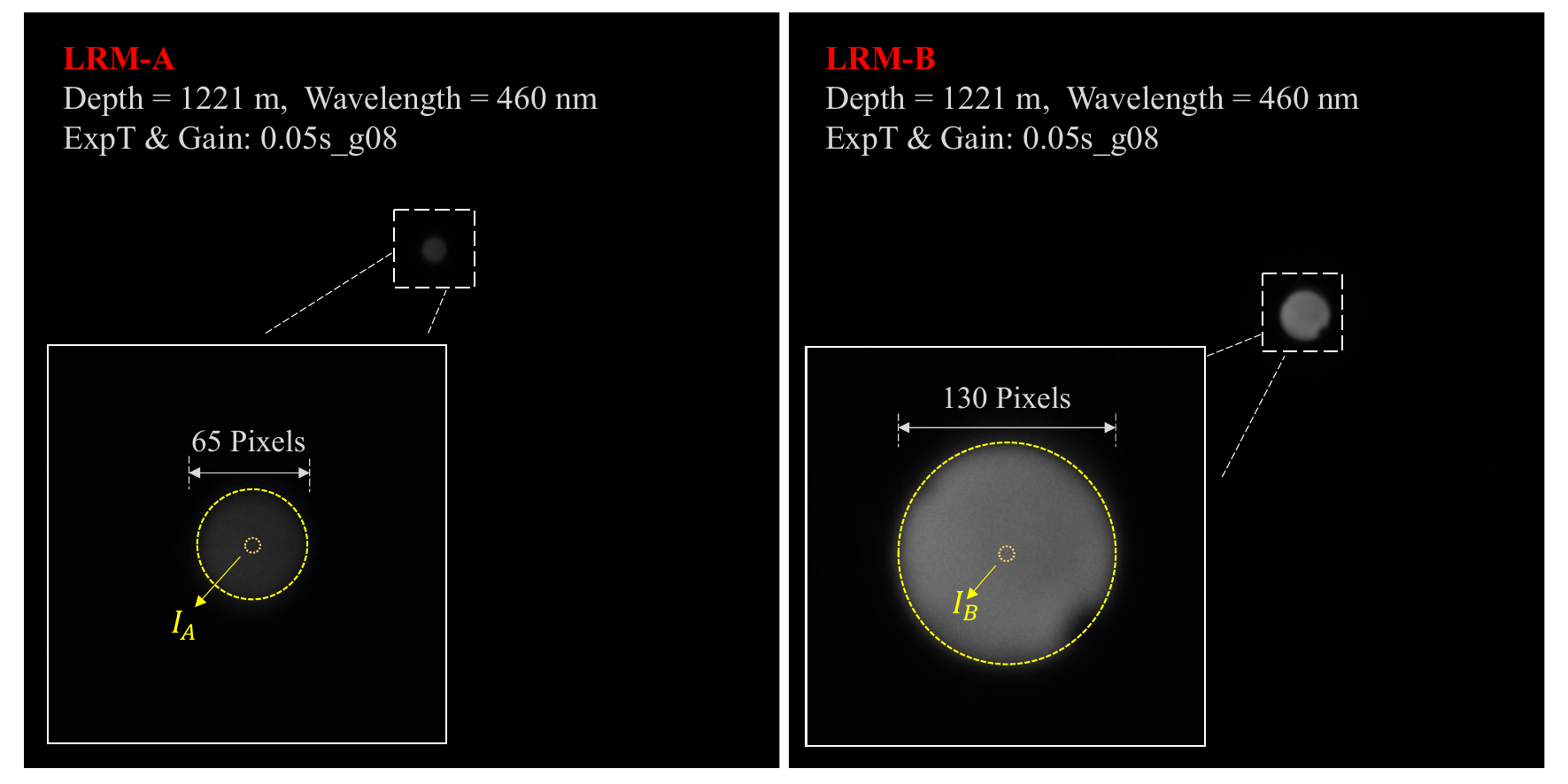}
% "\includegraphics" from the "graphicx" permits to crop (trim+clip)
% and rotate (angle) and image (and much more)
\caption{\label{fig image_cut} An illustration of image processing. The dashed yellow circle indicates the profile of the LEM on each images. The central region is selected for calculating the mean gray value of the centroid pixel.}
\end{figure}

% As shown in figure \ref{fig I_center}, $L_\mathrm{{A}}$ and $L_\mathrm{{B}}$ are the two measurement distances and $L_{A}\approx2L_{B}$.  are mean gray values of the center pixels of the LEM images took by Cam-A and Cam-B with the same exposure time and gain settings. They are proportional to the light intensity within the same solid angle but attenuated by two distances in water. Considering the imperfection of the LEM, we  The influence factor are canceled
%\subsection{Imaging processing for $I_{\mathrm{center}}$ method}
%这里，使用相对测量的方法的好处在于，可以消除一些系统不确定性，包括折射率变化导致的界面的光损耗，深海低温环境对于整个系统的影响，以及相机CMOS对于不同波长光的响应效率的影响等\ref{}。此外，利用这个方法，理论上仅仅需要Cam-A以及Cam-B按照相同的参数设置各拍摄一张照片，就可以实时测量出水的衰减长度，十分高效且准确。
For the image processing in the $I_{\mathrm{center}}$ method, we first filter out images with no saturated pixels. Given an 8-bit camera, the upper limit for gray values is 255, and our pre-calibration confirms a safe linear response region for gray values ranging from 5 to 240, as detailed in \ref{linearity_cal}. Within each selected image, we analyze the gray value distribution and locate the centroid of the LEM. Subsequently, we crop the image around this centroid to obtain a uniform $300\times300$ pixel size, as illustrated in Figure \ref{fig image_cut}. This cropping procedure ensures precise alignment of the LEM in each image and facilitates subsequent stacking analysis, thereby reducing statistical uncertainties.

\begin{figure}[!ht]
\centering % \begin{center}/\end{center} takes some additional vertical space
\includegraphics[width=.99\textwidth]{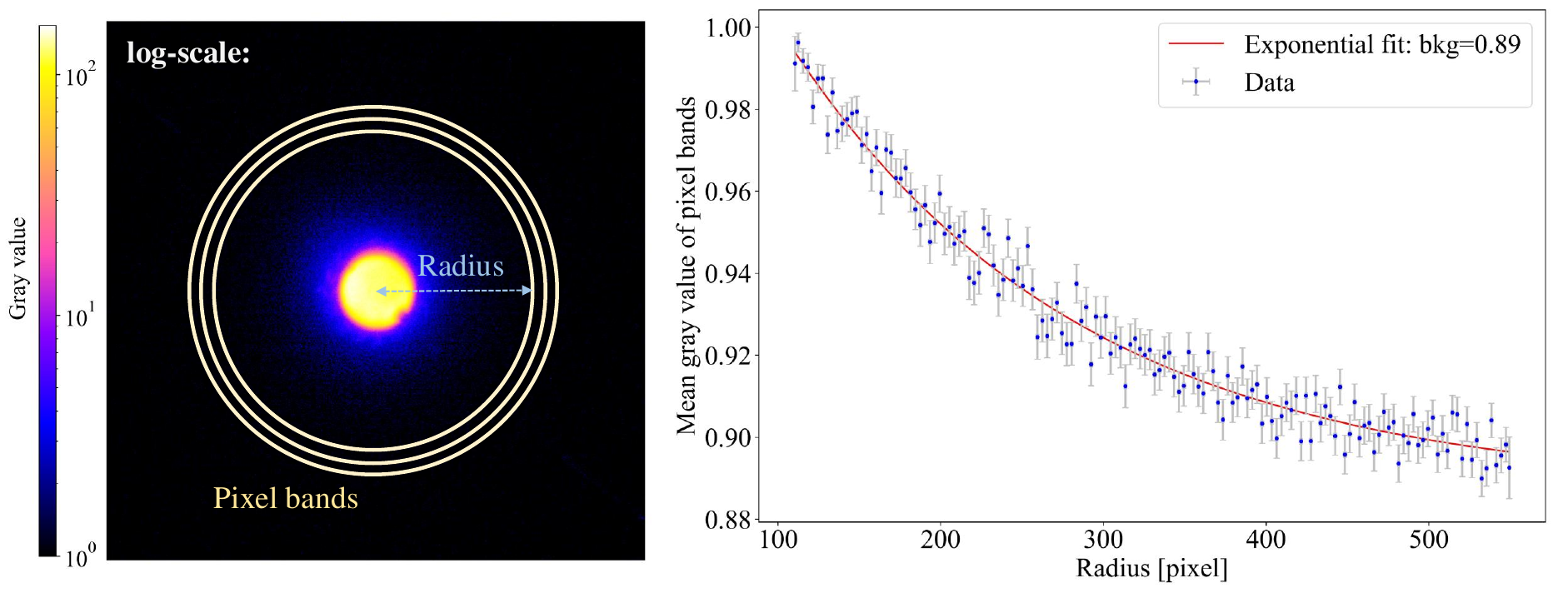}
% "\includegraphics" from the "graphicx" permits to crop (trim+clip)
% and rotate (angle) and image (and much more)
\caption{\label{fig bkg_fit} An illustration of the fitting method to obtain the image background. The left plot shows the gray value distribution of the image under the log scale, which can depict the weak scattered light. The right plot shows the mean gray value of each pixel band, and the image background can be obtained by an exponential fitting.}
\end{figure}

In the next step, we employ a fitting method to extract the intrinsic noise from CMOS pixels, as illustrated in Figure \ref{fig bkg_fit}. Since the scattered photons have a widely spread distribution on the image, we analyze the gray value distribution in the outer region of the image, which encompasses both scattered photons and the random background noise. To minimize uncertainties arising from the integer gray values near zero, we define pixel bands where all the pixels have an identical radius from the centroid and then calculate the mean gray value of each pixel band. The mean gray values of these pixel bands exhibit an exponential decrease as the radius increases, enabling us to extract the constant background contribution via fitting. Subsequently, we calculate the mean gray value of 100 pixels around the centroid and subtract the background contribution to obtain $I_\mathrm{{A}}$ and $I_\mathrm{{B}}$, which are then used to determine the attenuation length $\lambda_{\mathrm{att}}$.

% some other systematic uncertainties that commonly affect two cameras, such as light loss at the glass shell interface and the effect of low temperature in deep-sea environments, particularly the response efficiency of the camera CMOS.
In terms of uncertainty estimation in the $I_{\mathrm{center}}$ method, systematic uncertainties primarily stem from the experimental setup of measurement distances $L_A$ and $L_B$, as well as the calibration of the light source. Other systematic uncertainties, which equally impact both LRMs, such as photon loss at glass shell interfaces as discussed earlier, and the response efficiency of the camera's CMOS sensor under deep-sea temperatures, can be mitigated through the relative measurement strategy. Statistical uncertainty arises from the calculation of the mean gray value.

In summary, the $I_{\mathrm{center}}$ method demonstrates great robustness to measure attenuation length, even if Cam-A and Cam-B capture only a few images each. It can be applied for real-time optical calibration of the future TRIDENT project.

\subsection{Statistical $\chi^2$ test for $\lambda_{\mathrm{abs}}$ and $\lambda_{\mathrm{sca}}$ measurement}

To further obtain the $\lambda_{\mathrm{abs}}$ and $\lambda_{\mathrm{sca}}$, we established a statistical $\chi^2$ model to extract the best-fitting optical parameters by comparing the gray value distribution of the simulated images and the real images. We scanned the possible parameter space of $\lambda_{\mathrm{abs}}$, $\lambda_{\mathrm{sca}}$, $\left \langle \mathrm{cos}\theta \right \rangle$ and refractive index for all three wavelengths in Geant4 simulation \cite{Sim_paper}.
% To further obtain the $\lambda_{\mathrm{abs}}$ and $\lambda_{\mathrm{sca}}$ from the attenuation effect, we established a statistical $\chi^2$ model to extract the best-fitting results by comparing the gray value distribution of the simulated images and the real images. We generated a huge simulation data set by Geant4 to traverse through the phase space of optical parameters for all three wavelengths. 

The pre-processing steps for real images in this method are the same as those for the $I_{\mathrm{center}}$ method mentioned earlier, which include image cropping and background fitting. After these steps, we convert the 2D image into a 1D gray value distribution profile by retaining the mean gray value of each pixel band. A joint normalization is then applied to both Cam-A and Cam-B data using the same normalization factor, denoted as $\sum_\mathrm{{i}}G_\mathrm{{i}}$. Here, $G_\mathrm{{i}}$ represents the gray value of the $i$-th pixel on the Cam-A image. This normalization process ensures that the initial gray value ratio between the two cameras is preserved.

\begin{figure}[htbp]
\centering
\includegraphics[width=.7\textwidth]{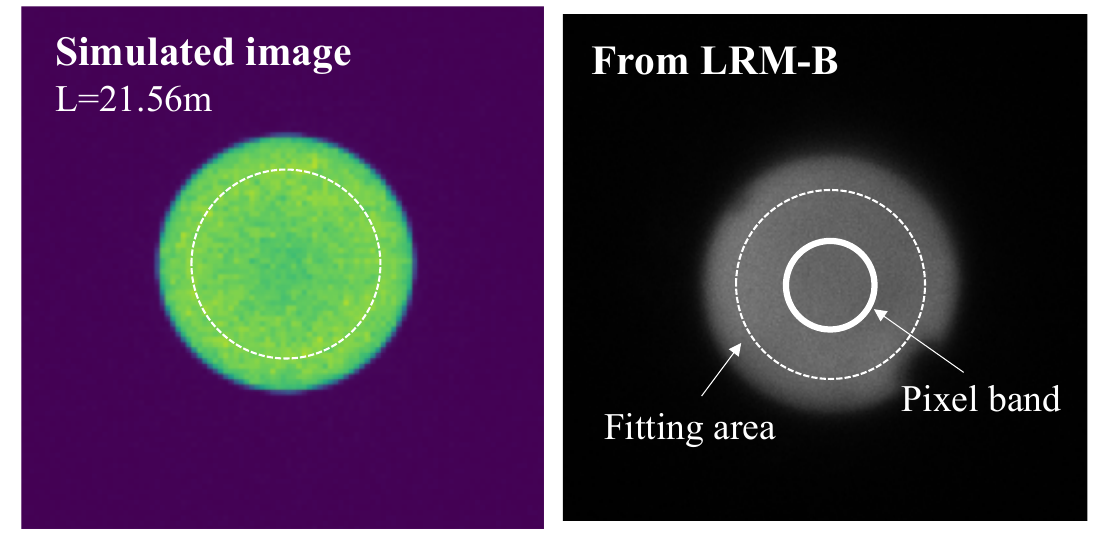}
\caption{\label{fig:camera_Chisq} A comparison of a simulated image example and a real images captured at 21.56 m, which are applied in statistical $\chi^2$ analysis method. The dashed white circle indicates the fitting area in this analysis.}
\end{figure}

For the simulation data processing, several steps are taken to incorporate the calibrated response characteristics of the cameras and the LEM. Firstly, we re-weight the emitted photons from the isotropic light source in the simulation based on a pre-calibrated emission profile that describes the nonuniformity of our LEM. Secondly, we apply a 2D-Gaussian Point Spread Function (PSF) to the simulated images to account for the potential defocusing effect of the cameras. The parameter in the PSF is treated as a nuisance parameter and is used in the global fitting process. Additionally, we consider the factor $I'_\mathrm{{0}}/I_\mathrm{{0}}$ during the normalization of the simulated images before converting them into 1D arrays.

The $\chi^2$ value is then calculated bin by bin using the following model:

\begin{equation}
    \chi^2 = \sum_{i=1}^{N}{ \frac{ [ M_\mathrm{i} - T_\mathrm{i} (1+\sum_{k=1}^{K}{\epsilon_\mathrm{{k}}})]^2}{\sigma_\mathrm{{Mi}}^2 + \sigma_\mathrm{{Ti}}^2} } + \sum_\mathrm{{k=1}}^{\mathrm{K}}{ \frac{{\epsilon_\mathrm{{k}}^2 }}{\sigma_\mathrm{{k}}^2}} .
\end{equation} 

Here, $M_\mathrm{{i}}$ represents the normalized mean gray value of the $i$-th pixel band in the real image, while $T_\mathrm{i}$ corresponds to the simulated value. $\sigma_\mathrm{{Mi}}$ and $\sigma_\mathrm{{Ti}}$ are uncorrelated uncertainties, such as statistical fluctuations in the measured gray value of the pixel bands and the simulated photon number in the $i$-th bin. $\epsilon_\mathrm{{k}}$ represents nuisance parameters added as penalty terms, which account for several uncorrelated systematic uncertainties including uncertainties in $L_\mathrm{{A}}$ and $L_\mathrm{{B}}$, the factor $I'_\mathrm{{0}}/I_\mathrm{{0}}$, and the parameter setting of the Gaussian PSF.
\begin{figure}[htbp]
\centering
\includegraphics[width=.7\textwidth]{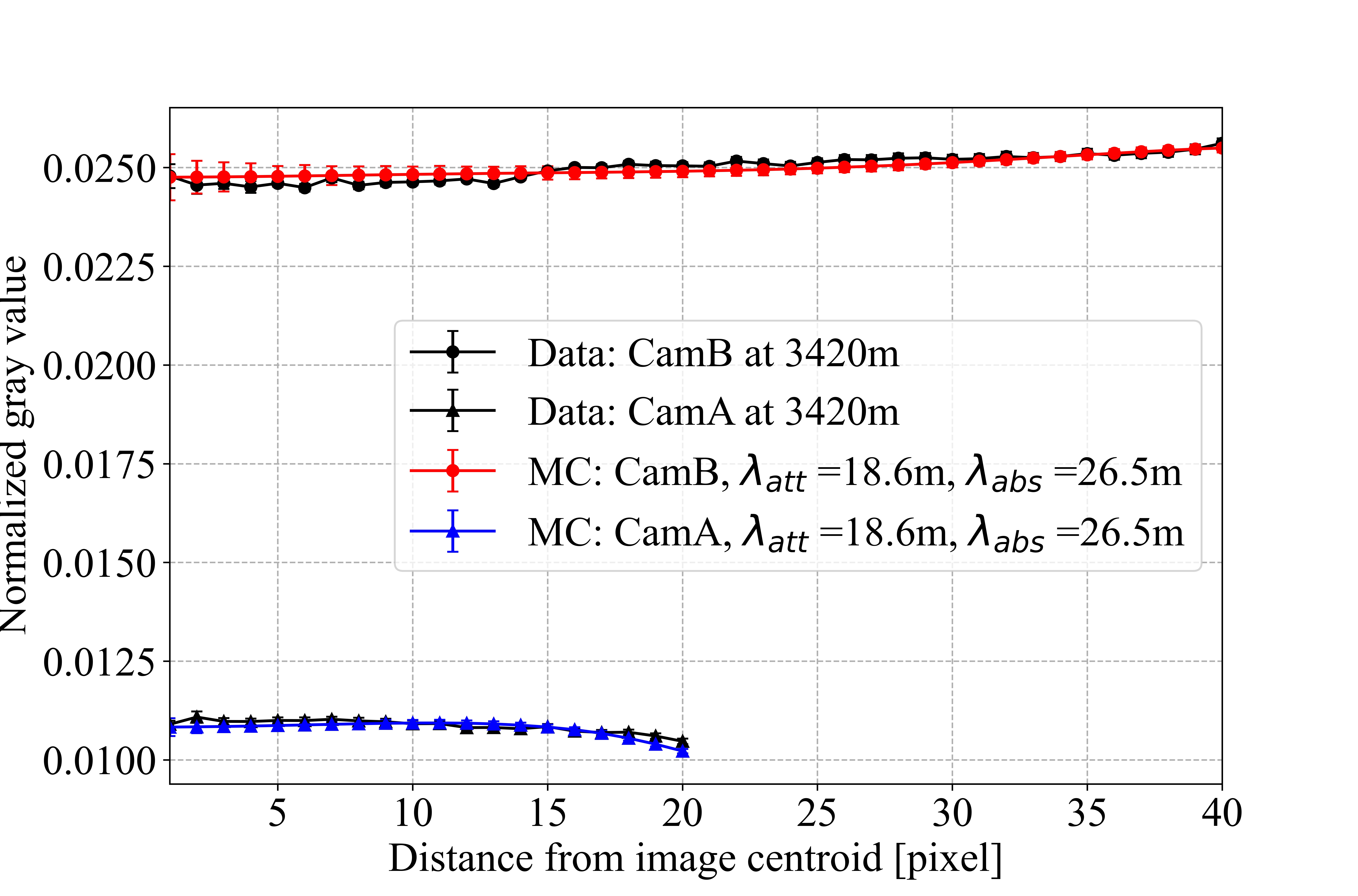}
\caption{\label{fig:camera_best_fit} The optical parameters with the best fit to the data recorded at 3420 m. The black points represent the data obtained from real images, while the red and blue points correspond to the simulated data.}
\end{figure}

By minimizing $\chi^2$ within the simulation phase space, we can determine the values of $\lambda_{\mathrm{abs}}$ and $\lambda_{\mathrm{sca}}$ separately, as depicted in Figure \ref{fig:camera_best_fit}. However, distinguishing between Mie and Rayleigh scattering from the overall effect of $\lambda_{\mathrm{sca}}$ remains challenging. A more detailed distinction can be achievable through $\chi^2$ fitting of the full image area, encompassing the weak scattering light illustrated in Figure \ref{fig bkg_fit}. This aspect is currently under development.

\section{Calibration of the camera system}
\label{sec:Calibration}

%uniformity and low-temperature verification (non-uniform emitter, modulation code for chi-square analysis, ,)------refrigerator, drag water tank
%sensitivity (able to find the att and abs length in in-situ experiment; verification of the measurement strategy) ------sink
%algorithm and code (hough circle, )-----Zhiyuan second floor
%capacity (verification of hardware[control system design, data transmission and storage, camera])-----Zhiyuan second floor, drag water tank
%adjust the exposure time and gain to make sure the measurement is valid(neither saturated nor too weak, focal length)-----Zhiyuan second floor, drag water tank
%linear response -----self calibration
% The choice of 

% , also exposure time and gain settings of camera system depends on the 

% camera technical parameters is determined and the sensitivity of the instrumentation to measure the attenuation length is verified. Non-uniformity of the light source, linear response of the camera, low temperature influence and focal length change due to the water-glass interface are all considered in detail.

To calibrate the performance of the camera system, several experiments were conducted to test both the hardware and measurement methods, ensuring their suitability for the deep-sea experiment.

\subsection{Linear response calibration}
\label{linearity_cal}

One important aspect is to verify the linear response of the CMOS sensor of the cameras. In this test, a steady light source was installed in a darkroom, and images were captured with varying exposure times ranging from 1 ms to 350 ms while the gain setting on the cameras remained the same. Since the exposure time is directly proportional to the light intensity, the camera's response can be evaluated by analyzing the gray values of identical pixels of the captured images. 

\begin{figure}[htbp]
\centering % \begin{center}/\end{center} takes some additional vertical space
\includegraphics[width=0.95\textwidth]{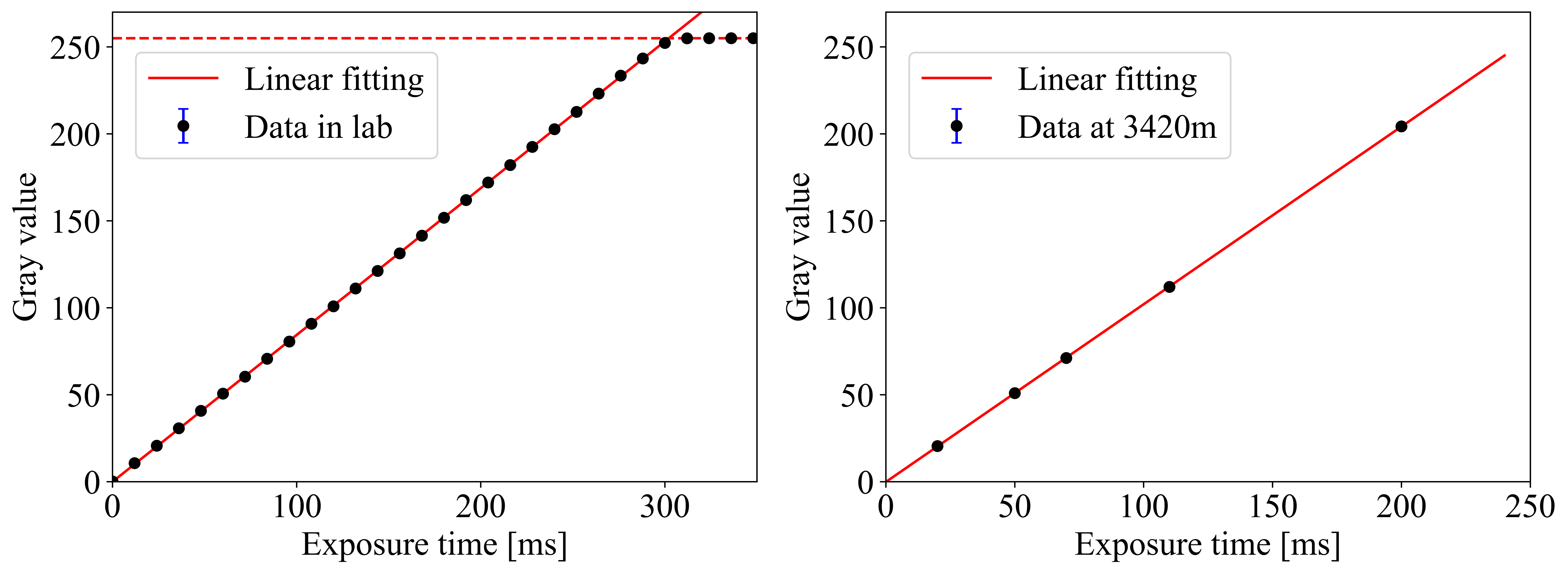}
% "\includegraphics" from the "graphicx" permits to crop (trim+clip)
% and rotate (angle) and image (and much more)
\caption{\label{fig_linear_response} The left panel shows laboratory test results, while the right panel depicts self-calibration data from the actual deep-sea experiment at a depth of 3420 m. Both demonstrate the camera's stable linear response with a goodness of fit larger than 0.99.}
\end{figure}

The test results, as shown in Figure \ref{fig_linear_response}, demonstrate a stable linear response within the operating range of the cameras before reaching the gray value saturation. This figure also illustrates the in-situ self-calibration performed during the deep-sea experiment, further ensuring the consequent analysis. 

%In the likelihood, we need to compare the simulated data and experimental data, which unit are number of photons and gray value, respectively. Validating the linear response is of importance in the likelihood method since we need to compare these two curves. In the images, the gray value of the image is related to the exposure time, the gain and the environment it works.
% Besides, the number of photons is proportional to the exposure time. We use the data with the gain equal to 2 at the depth of 3399m to investigate the relation between the gray value and the exposure time. The linear response calibration is shown in Figure~\ref{fig_linear_response}, where the x axis is the exposure time which is related to the number of photons and the y axis is gray value of the image. Figure~\ref{fig_linear_response} shows a good linear response between the gray value and the number of photons, which verify our method in the likelihood make sense. 

% \subsection{Nonuniformity Calibration of Light Source}
% Photos of the two sides of the light emitter module are obtained in the refrigeration laboratory. Two sides here points to the two hemispheres directly facing the light receiver module A and the light receiver module B respectively. The photos are taken in the light shading environment with the distance of approximately five meters, so that sufficient clarity and sharpness are guaranteed to observe the non-uniformity of the light source. 

%pics here

%The photos above carry the detailed information of the light intensity distribution of the light emitter module. A array of weights can be generated to calibrate the histogram outcome of simulated light intensity distribution. This operation refines the simulation results, which enables the calculation of the measurement results along with the confidence interval from Chi-square analysis. 

\subsection{Low-temperature calibration}
% The camera and its control board were tested at temperatures ranging from $2~4\pm 0.5^{\circ}\mathrm{C}$, while the LEM was tested at $2.1\pm 0.5^{\circ}\mathrm{C}$.

\begin{figure}[!ht]
\centering % \begin{center}/\end{center} takes some additional vertical space
\includegraphics[width=1\linewidth]{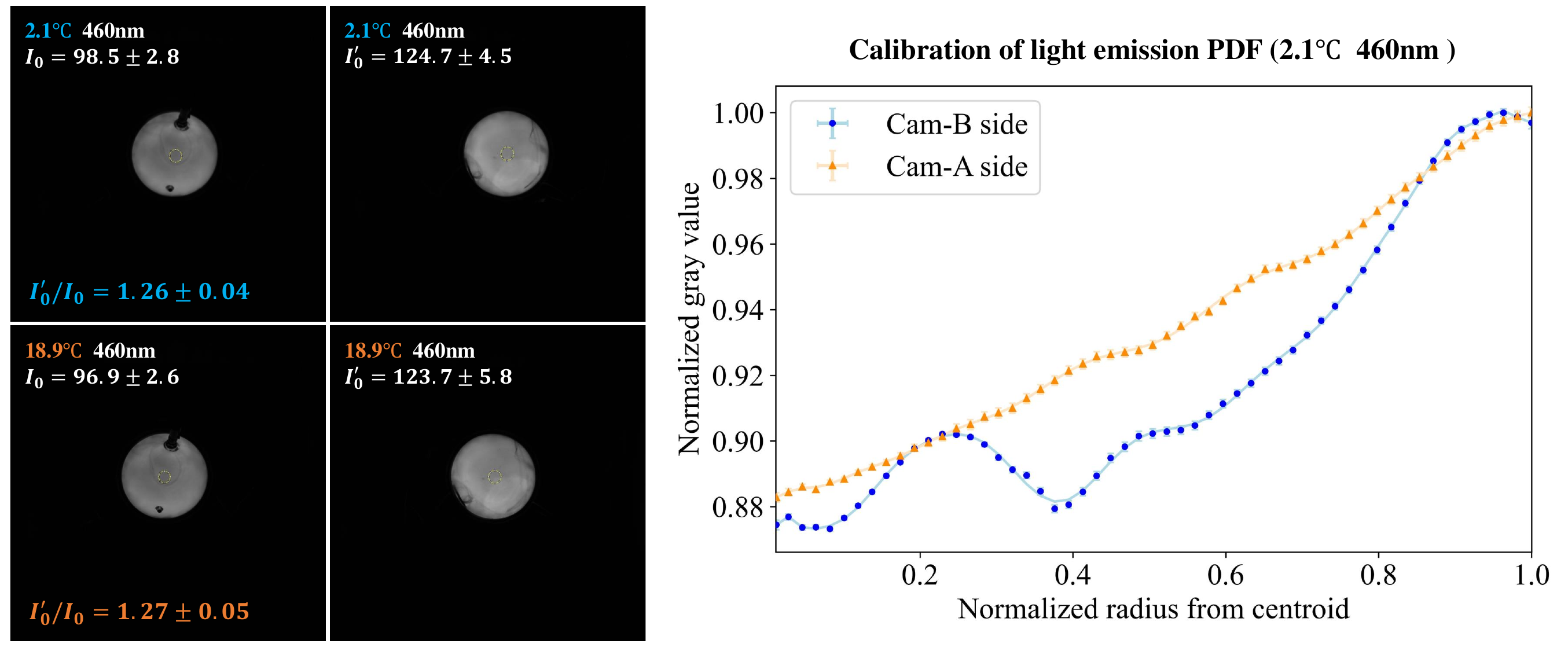}
% "\includegraphics" from the "graphicx" permits to crop (trim+clip)
% and rotate (angle) and image (and much more)
\caption{\label{low_tem_cali} The results from low-temperature calibration. The left plot shows the calibration results of $I_\mathrm{{0}}'/I_\mathrm{{0}}$ at two different temperatures at 460 nm. The right panel shows the normalized light emission profile of the LEM. }
\end{figure}

To ensure expected performance in deep-sea environments, we conducted a temperature-controlled experiment to calibrate the camera system together with the LEM.
As previously mentioned, $I_\mathrm{{0}}'/I_\mathrm{{0}}$ is a critical input parameter in the $I_{\mathrm{center}}$ method, encompassing the influence of both the emission non-uniformity of the LEM and the camera response. To evaluate this parameter's sensitivity to temperature variations, we conducted a test where we captured images under identical experimental settings at two temperatures: $2.1\pm 0.5^{\circ}\mathrm{C}$ and $18.9\pm 0.5^{\circ}\mathrm{C}$. We then compared the measured $I_\mathrm{{0}}'/I_\mathrm{{0}}$ values and estimated the potential uncertainty. The results indicate that the low-temperature effect on system performance is negligible, with an influence of less than $1\%$ for all three wavelengths. Calibration results at 460 nm, for example, are depicted in Figure \ref{low_tem_cali}.

Additionally, in this experiment, we accurately characterized the LEM's light emission profile, which accounts for the non-uniformity of light intensity emitted from different latitudes in both hemispheres. This profile is influenced by factors such as 3D printing tolerances, shadows of wires, and internal LED positioning. This calibrated profile served as an input for subsequent simulations and analyses to ensure reliable data interpretation.

% \begin{table}\centering[htpb]
% \renewcommand{\arraystretch}{1.5}
% %\setlength{\arrayrulewidth}{0.15mm}
% %\setlength{\doublerulesep}{0.55mm}
% \centering
%     \begin{tabular}{|c|c|c|}
%       \hline
%       \textbf{Wavelength} & \textbf{$I_{0}'/I_{0}$ at $2.1^{\circ}C$ } &  \textbf{$I_{0}'/I_{0}$ at $18.9^{\circ}C$}\\ 
%       \hline
%       460 nm & $1.27\pm 0.04$ & $1.29\pm 0.04$\\ 
%       \hline
%       525 nm & $1.23\pm 0.04$ & $1.25\pm 0.03$\\
%       \hline
%       405 nm & $1.04\pm 0.03$ & $1.06\pm 0.03$\\
%       \hline
%     \end{tabular}
%     \caption{Calibration results of $I_{0}'/I_{0}$ for three wavelengths.}
%     \label{tab:cali-results of I'/I}
% \end{table}

%The temperature data obtained in the in-situ measurement shows that the sea water with the depth of three thousands kilometers is around 1.5 $^{\circ}C$.

%This ratio along with the in-homogeneity calibration data will be used for interpreting the difference between the simulation and the deep-sea experiment data, enabling more elaborate likelihood analysis to be conducted. 

% \begin{figure}[!ht]
% \centering % \begin{center}/\end{center} takes some additional vertical space
% \includegraphics[width=.7\linewidth]{camera_figures/lightsource_1d_cali.jpg}
% % "\includegraphics" from the "graphicx" permits to crop (trim+clip)
% % and rotate (angle) and image (and much more)
% \caption{\label{lightsource_1d_cali} The calibration results from low temperature calibration.}
% \end{figure}

\subsection{Long-distance test in air}
\label{Long-distance test in air}

To ensure the suitability of the camera's focal length and viewing angle for long-distance measurements in the deep sea, we conducted tests in a dark environment with varying distances ranging from $5.2 \pm 0.1$ m to $41.2 \pm 0.1$ m. For each wavelength of the LEM, we adjusted its initial light intensity and selected specific camera exposure time and gain configurations to compensate for the wavelength-dependent attenuation effect underwater.

\begin{figure}[!ht]
\centering % \begin{center}/\end{center} takes some additional vertical space
\includegraphics[width=.9\textwidth]{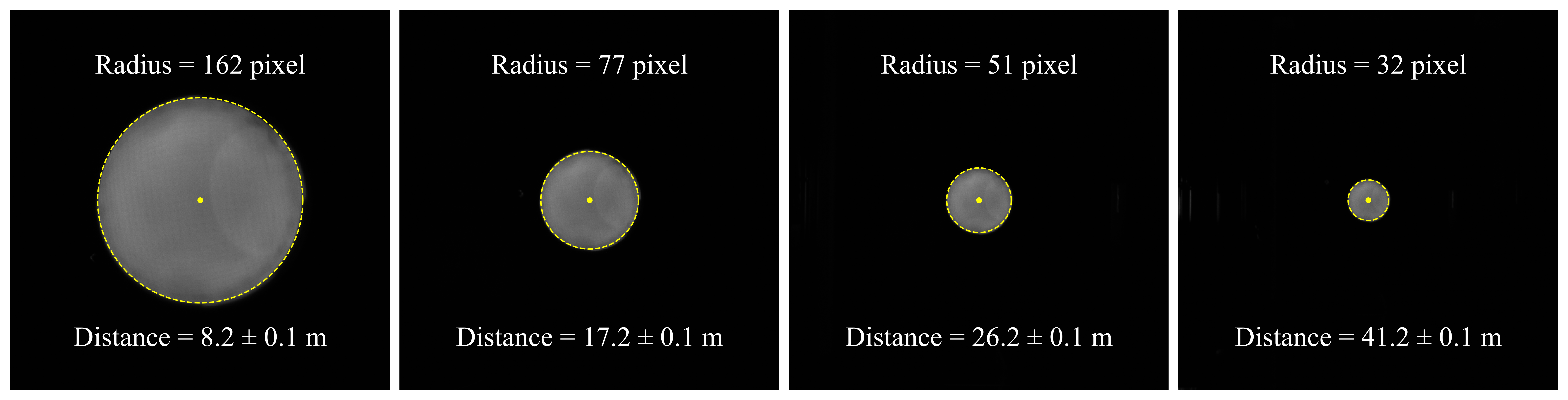}
\includegraphics[width=.6\textwidth]{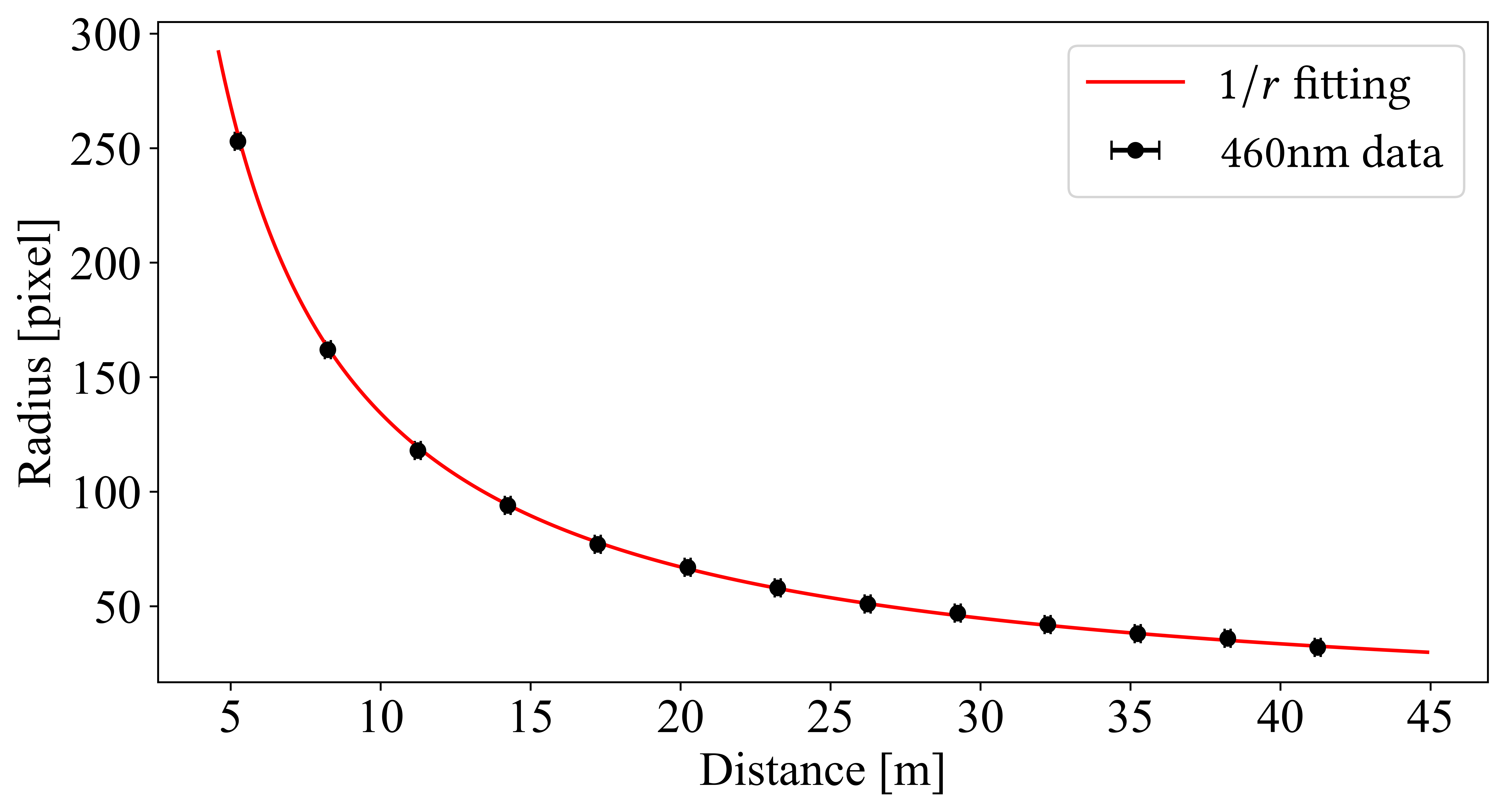}

% "\includegraphics" from the "graphicx" permits to crop (trim+clip)
% and rotate (angle) and image (and much more)
\caption{\label{fig_hough_air} The upper panel shows example images captured at four typical distances in the air, with dashed yellow circles indicating the radius obtained by the Hough-Circle algorithm. In the lower panel, the fitting result of the LEM radius varying with distances confirms the assumption of a pin-hole camera.}
\end{figure}

Additionally, we analyzed the recorded images at varying distances and applied the Hough-Circle algorithm to extract the LEM radius on the images, aiming to test potential distortion. The results, as illustrated in Figure \ref{fig_hough_air}, showed that the radius of the LEM in the images followed a $1/L$ decrease as the distance changed, confirming the assumption of a pin-hole camera.

\begin{figure}[htbp] 
  \renewcommand\figurename{\textbf{Figure.}}
    %\begin{subfigure}[htbp]{0.95\textwidth}
    \centering
    \includegraphics[width=0.49\linewidth]{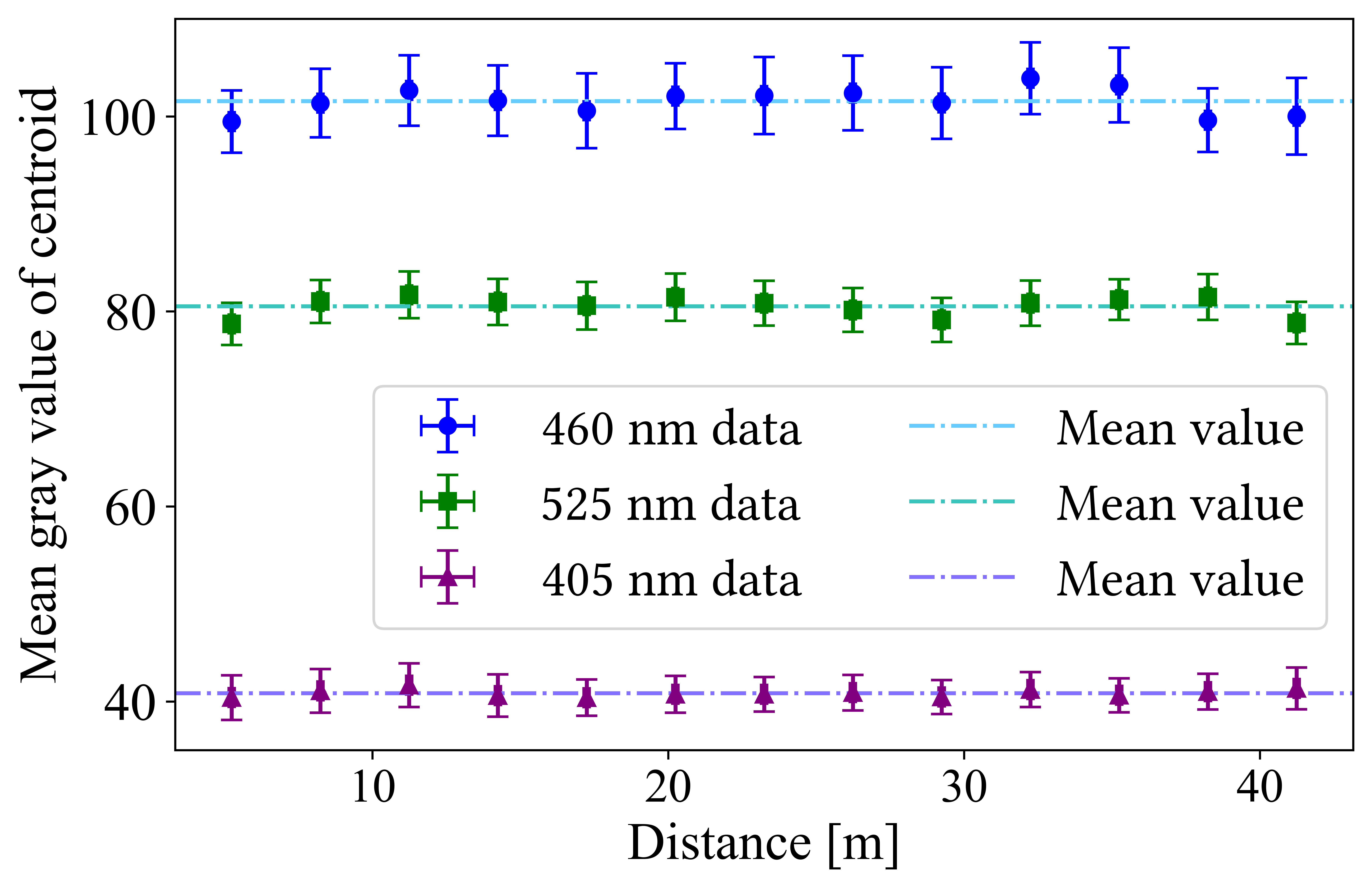}
    \includegraphics[width=0.49\linewidth]{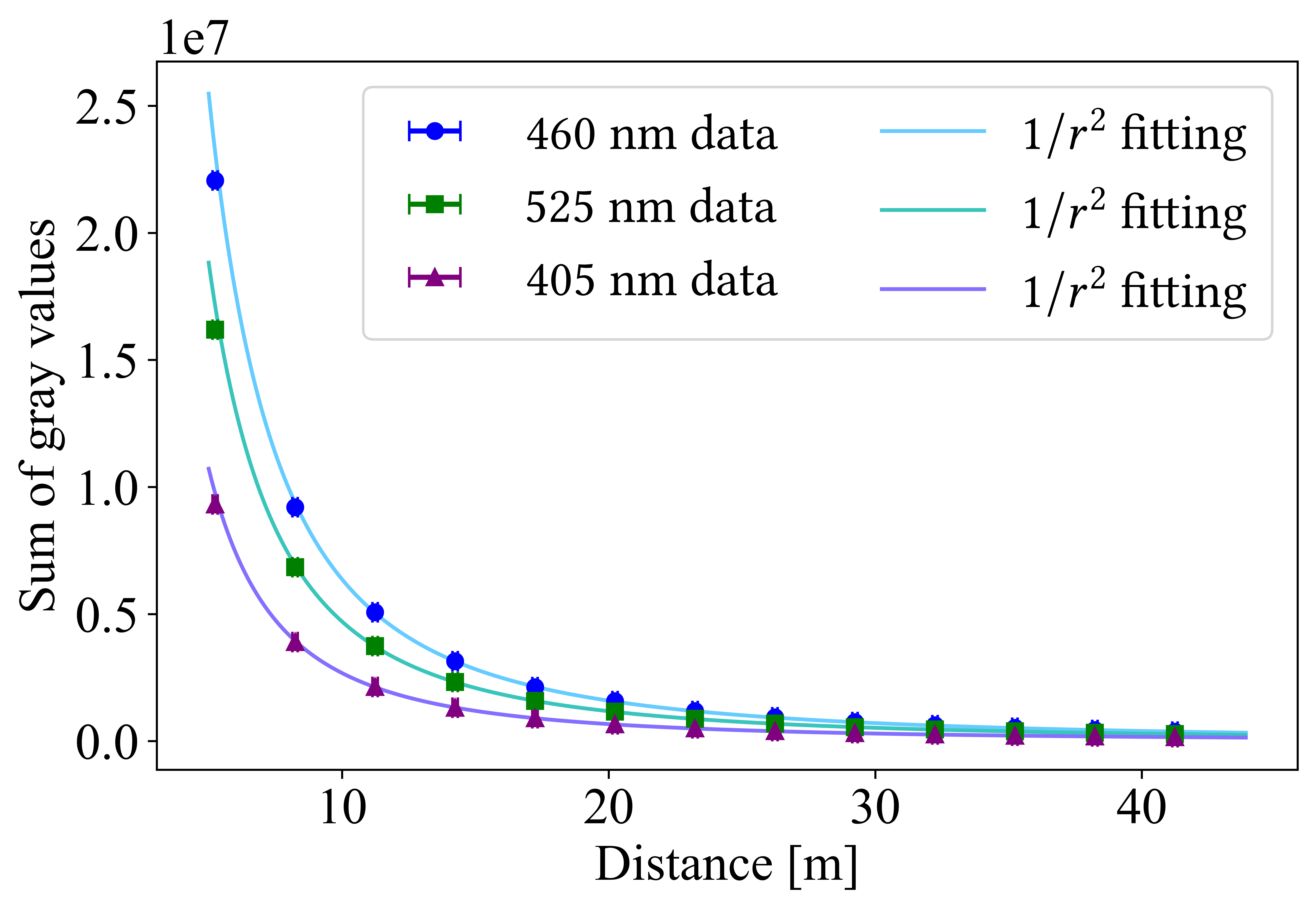}    

    \caption{The left panel shows the test results of the $I_{\mathrm{center}}$ method conducted in the air for all three wavelengths. The mean gray values of the LEM centroid at varying distances remain almost constant, reflecting the negligible attenuation effect. In the right panel, the sum of gray values in the image follows a $1/L^2$ exponential decrease at different measurement distances in the air.} 
    %\end{subfigure}
\label{air_test}
\end{figure}

Furthermore, this experiment can also serve as a control test for the $I_{\mathrm{center}}$ method. According to Formula \ref{I_center}, the mean gray value of the centroid pixel should remain nearly constant as the distances change because the attenuation effect of air is negligible at this distance level. This was confirmed by the results, as depicted in Figure \ref{air_test}. Furthermore, we verified that the sum of all gray values followed the expected $1/L^2$ decrease, as determined by Formula \ref{TW model}.

\subsection{Water tank experiment}
%\subsubsection{Setup}
To simulate the underwater experiment in the laboratory, we designed a custom water tank and conducted a scaled-down experiment to test the camera system and measure the attenuation length of tap water using the $I_{\mathrm{center}}$ method.

\begin{figure}[!ht]
\centering % \begin{center}/\end{center} takes some additional vertical space
\includegraphics[width=.99\textwidth]{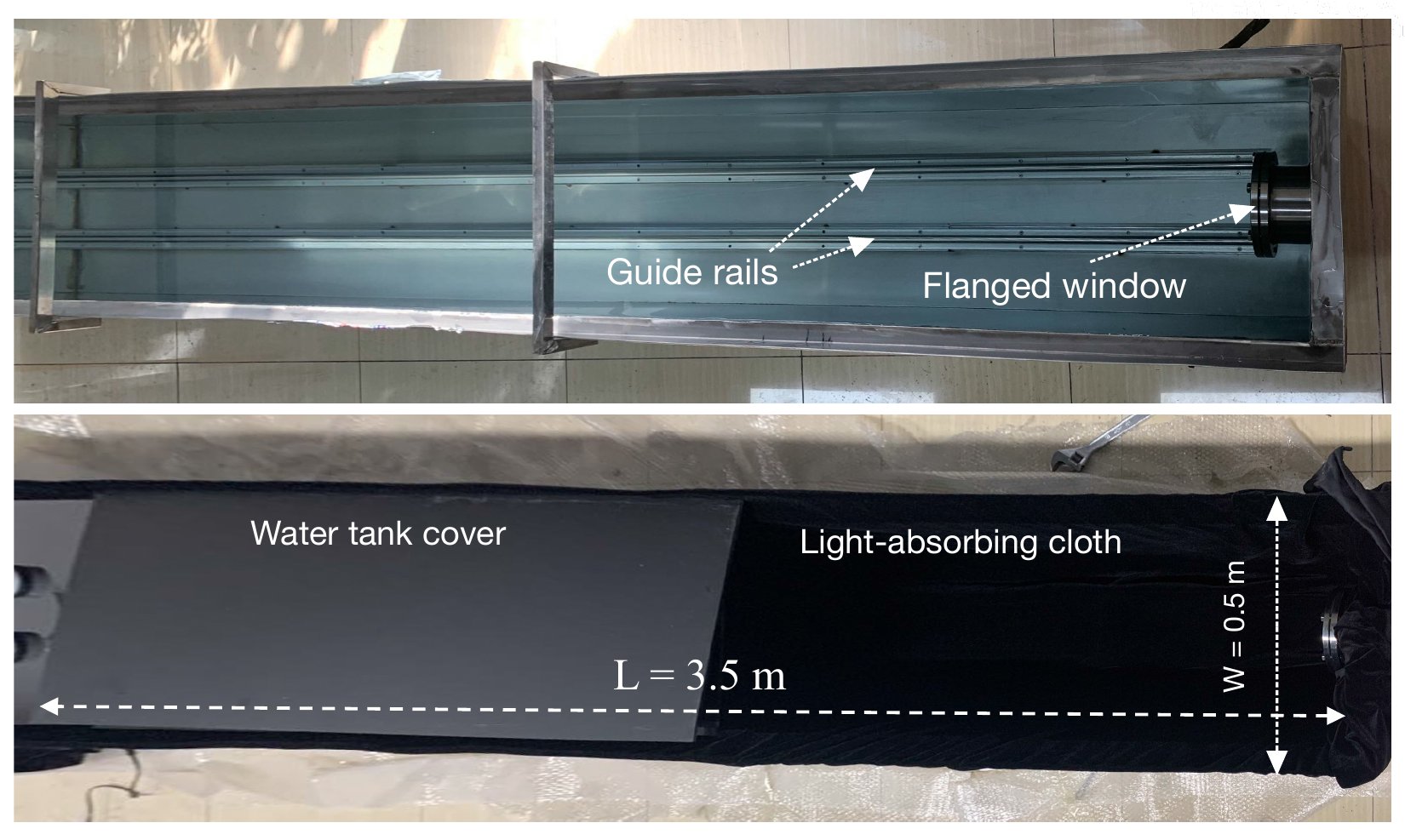}
% "\includegraphics" from the "graphicx" permits to crop (trim+clip)
% and rotate (angle) and image (and much more)
\caption{\label{fig_tank} Experimental setup of the water tank. The upper panel showcases the internal design, including the flanged window and the guide rails. Meanwhile, the lower image shows the setup with the light-absorbing cloth and a cover on top.}
\end{figure}

As depicted in Figure \ref{fig_tank}, the stainless steel water tank was crafted with dimensions of 3.5 m in length, 0.5 m in width, and 0.5 m in height. It featured a flanged window welded on one side and two parallel guide rails on the bottom. The flange, with a diameter of 25 cm, housed an organic glass window to accommodate the camera inside the tube and isolate it from the water. The guide rails, measuring $320 \pm 1$ cm in length, were equipped with a precise length scale with 1 cm accuracy to facilitate the movement of a light source platform along the rails. The design with two guide rails helped minimize movement uncertainties in other dimensions during the sliding operation.

For the experiment, we used a miniature light source comprising a single LED embedded in a 2.5 cm Teflon ball, which served as both a light diffuser and a waterproofing outer shell. During the experiment, we placed a large light-absorbing cloth on the inner surface and added three black covers to the top of the tank, all of which were immersed in water. This setup aimed to minimize light reflection on the internal stainless steel surfaces and the water-air interface, with the reflection rate measured to be less than 5\%.

\begin{figure}[!ht]
\centering
\includegraphics[width=1\linewidth]{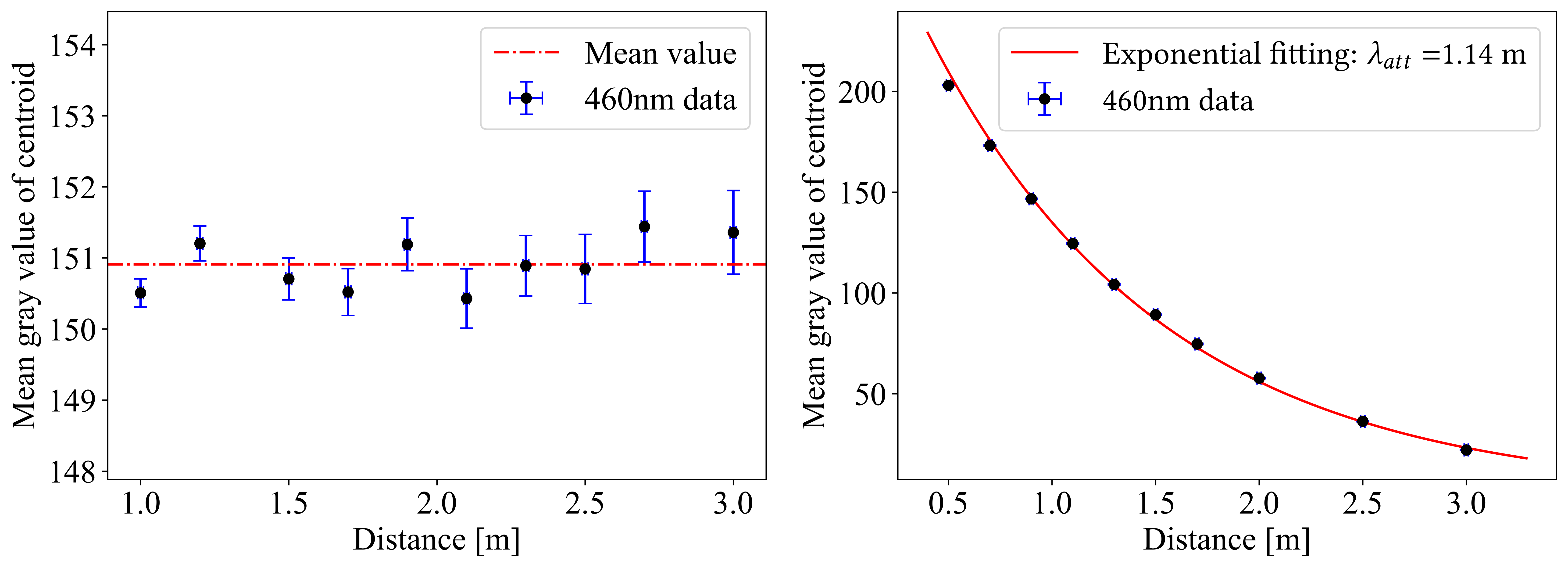}
\caption{\label{fig_tank_results} Test results of the $I_{\mathrm{center}}$ method in the water tank experiment. The left panel shows the nearly constant gray values of the light source centroid obtained at varying distances in the air. The right panel illustrates the fitting result of the exponential decay of the mean gray value of the centroid in the water.}
\end{figure}
During the experiment, the camera first recorded images of the light source at distances ranging from $100 \pm 1 $ cm to $300 \pm 1$ cm without water. Then the measurements were repeated after filling the tank with tap water. The compared results, shown in Figure \ref{fig_tank_results}, indicated that the mean gray value of the light source centroid remained nearly constant in the air, consistent with the results of the previous long-distance test. In water, the decay of the mean gray value followed an exponential law, as described in the $I_{\mathrm{center}}$ method, and the attenuation length was fitted.

This experiment successfully verified the $I_{\mathrm{center}}$ method in water, and it also demonstrated the water tank as a useful facility for laboratory underwater calibration of optical modules in the future.

%这里要说明 上板是经过哑光处理，并且浸没在水中的，这样的话可以消除空气和水界面处的光反射。

%从视角而言，2.5cm的特氟龙光源小球相当于是 1.2 m和2.4m左右的viewing angle

%我们利用水槽进行对比实验，先在空气中进行50-300 cm-的测量，之后注满自来水，

% 这里的问题还是，标准差要不要再除以 根号N
% 图片名带有 _overN 的表示除了 根号N

% \begin{figure}[!ht]
% \centering % \begin{center}/\end{center} takes some additional vertical space
% \includegraphics[width=.4\textwidth]{camera_figures/blue_test.png}
% % "\includegraphics" from the "graphicx" permits to crop (trim+clip)
% % and rotate (angle) and image (and much more)
% \caption{\label{fig_blue_test} Here is the picture of the sink.}
% \end{figure}

%\subsubsection{Distance-Brightness Relationship in tap water}

\subsection{Focal length calibration in water}
\label{towing tank}

To account for the optical refraction caused by the glass shell's curvature and the refractive index difference between the inside and outside of the shell, we conducted an experiment in a large ship model towing tank located on the Minahng campus of Shanghai Jiao Tong University to re-calibrate the focal length of our cameras for underwater imaging.

\begin{figure}[!ht]
\centering % \begin{center}/\end{center} takes some additional vertical space
\includegraphics[width=1 \linewidth]{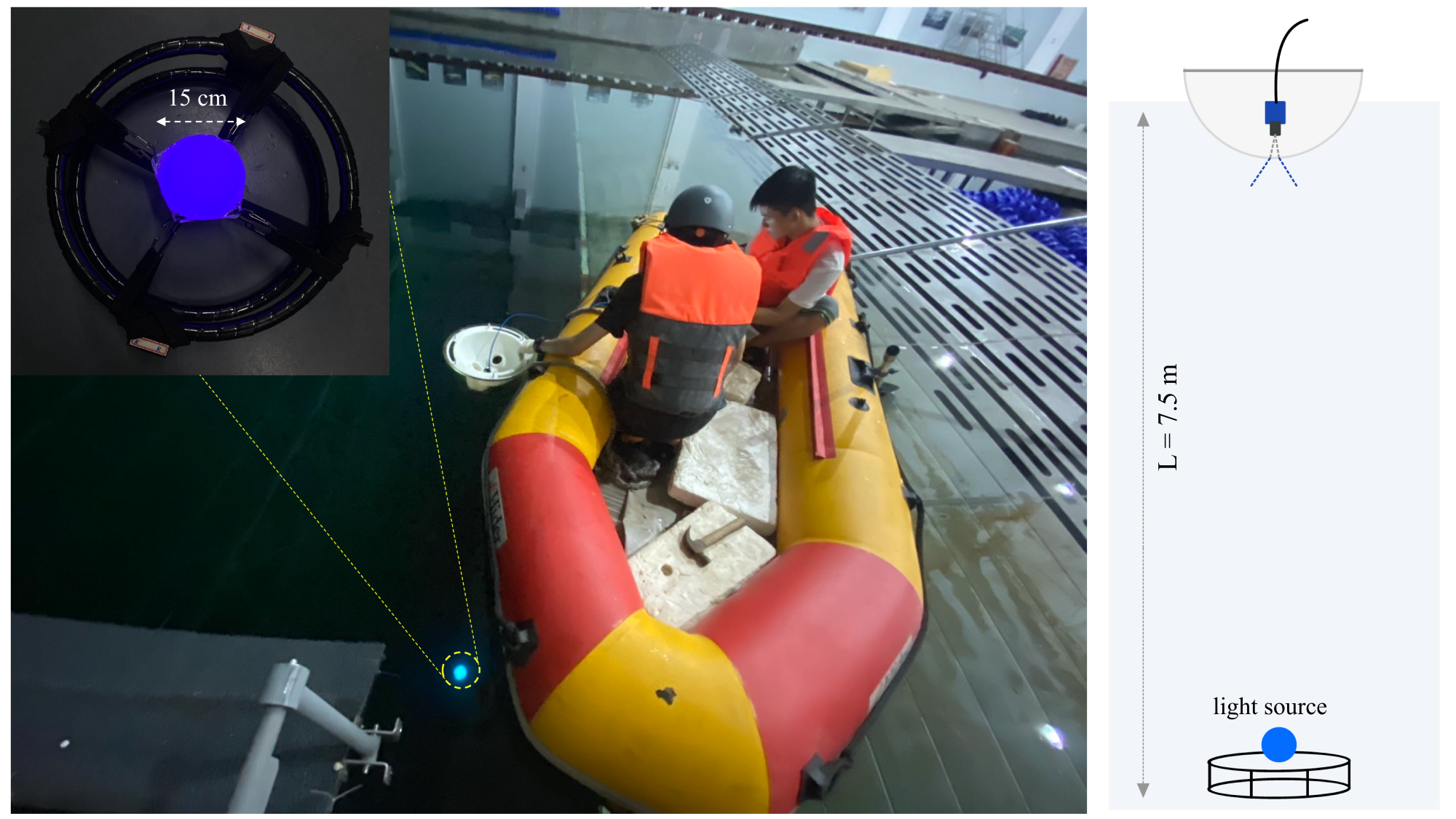}
% "\includegraphics" from the "graphicx" permits to crop (trim+clip)
% and rotate (angle) and image (and much more)
\caption{\label{towing_tank} An illustration of the towing tank experiment for testing the additional optical refraction effect. A scaled-down waterproof light source was deployed to the bottom of the towing tank to help adjust the camera's focal length.}
\end{figure}

The towing tank has a depth of 7.5 meters, a length of 300 meters, and a width of 15 meters, providing an ideal experimental environment. During the calibration process, we deployed a waterproof light source with a 15 cm diameter to the bottom of the towing tank. The camera captured images through the glass shell on the water's surface, as depicted in Figure \ref{towing_tank}. By adjusting the camera's focal length in this setup, we aimed to ensure clear image recording under underwater conditions.

The viewing angle occupied by the miniature light source at a depth of 7.5 m was approximately equivalent to that of the LEM with a diameter of 43 cm placed at a distance of 21.6 m, corresponding to the experimental setup of T-REX. Additionally, the brightness of the waterproof light source could be controlled remotely, allowing us to simulate various water conditions and adjust the camera's exposure time and gain settings accordingly.

Given that the water's refractive index slightly differs from actual seawater and considering the different depths of fields for Cam-A and Cam-B in T-REX, which could also mildly affect the imaging process, we introduced a PSF for our data analysis, as discussed earlier. In the future, autofocus can be considered a significant update for the camera system.

\section{Conclusion and outlook}
\label{conclusion_outlook}
In this paper, we present a costum-designed camera system tailored for real-time optical calibration in water-based neutrino telescopes, addressing the potential time-varying or non-uniform optical parameters due to the dynamic water. This system has been successfully demonstrated in TRIDENT's Pathfinder experiment.

We discuss existing light propagation models used in previous experiments and have introduced a modified version that accurately describes photon propagation in a less-scattering water medium, originating from a spherically symmetrical light source. This refined model accounts for the combined effects of scattering and absorption using only canonical optical parameters.

We then introduced two analysis methods: the $I_{\mathrm{center}}$ method for rapid measurement of the attenuation length ($\lambda_{\mathrm{att}}$) and a statistical $\chi^2$ test for decoding the absorption length ($\lambda_{\mathrm{abs}}$) and scattering length ($\lambda_{\mathrm{sca}}$). A comprehensive laboratory-based calibration procedure was conducted to account for potential factors influencing the camera system's performance.

For future application, the camera system's compact size enables easy installation within the glass pressure vessel, and its remote control capability facilitates image capture under varying exposure times and gain settings to accommodate different water conditions. Efficient data transmission and image processing algorithms make it well-suited for real-time optical calibration of water-based neutrino telescopes. Our future endeavors include system upgrades, such as autofocus for cameras, integration into new calibration modules, and the development of a more comprehensive calibration method using a fully-functional water tank. 

\section{Acknowledgments}
% \acknowledgments

% This is the most common positions for acknowledgments. A macro is
% available to maintain the same layout and spelling of the heading.
We extend our sincere gratitude to Jun Guo for his insightful comments and discussions, which significantly contributed to the enhancement of this paper. Our thanks also go to Woosik Kang, Yangjie Su, Xinhai Xie, Kun Xu, Qi Sun, Li Song, Xiaoliang Zhang, and Liangyu Wang for their valuable discussions and suggestions on CMOS camera technology, photon propagation models, auto-control systems, and 3D printing supports. Additionally, we are grateful to the State Key Laboratory of Ocean Engineering at Shanghai Jiao Tong University for granting us permission to use the ship towing tank for camera calibration.

This work has received support from the Ministry of Science and Technology of China under Grant No. 2022YFA1605500, the Office of Science and Technology of the Shanghai Municipal Government under Grant No.22JC1410100, Double First Class startup fund and the Foresight grants No. 21X010202013 and No. 21X010200816 provided by Shanghai Jiao Tong University.

\appendix

% 传感器部分
\section{Supplementary sensor data analysis}

\label{App A}

\begin{figure}
\centering
\includegraphics[width=1\linewidth]{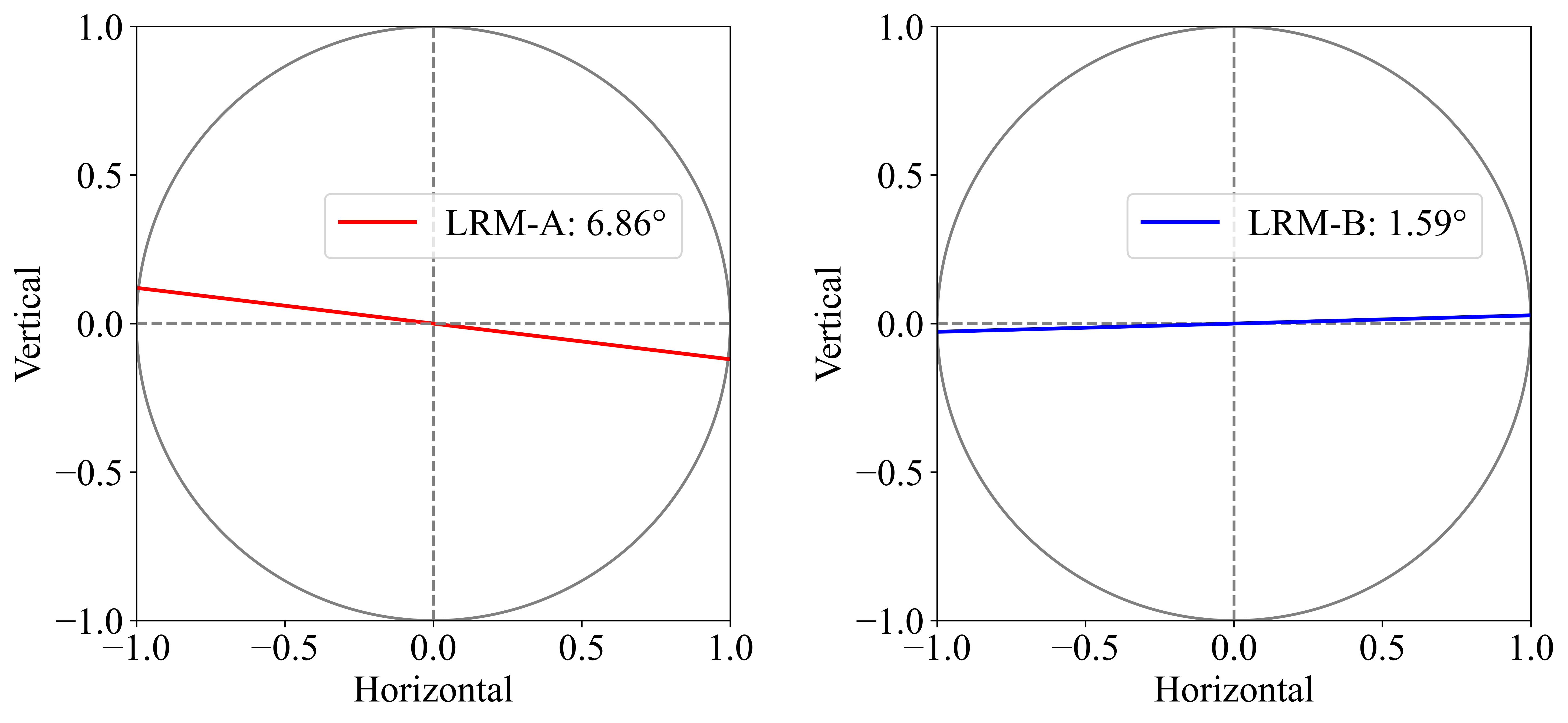}
\caption{\label{fig_sensor_slope} An example of the real-time gestures performed by two LRMs during the T-REX deployment. It showcases the angular attitude reconstruction of LRM-A and LRM-B, presenting the existing but acceptable tilt.}
\end{figure}

During the T-REX deployment, ocean currents can affect the apparatus's orientation, potentially causing the LRMs to tilt and deviate from the vertical direction. This deviation reflects on the LEM shifting from the center of the camera's viewing angle. To monitor this influence, accelerometers were installed inside the two LRMs, as introduced above. By analyzing the acceleration variations in three directions, the angle of inclination of the LRM from the vertical direction can be reconstructed, as shown in Figure~\ref{fig_sensor_slope}.

\begin{figure}
\centering
\includegraphics[width=1\linewidth]{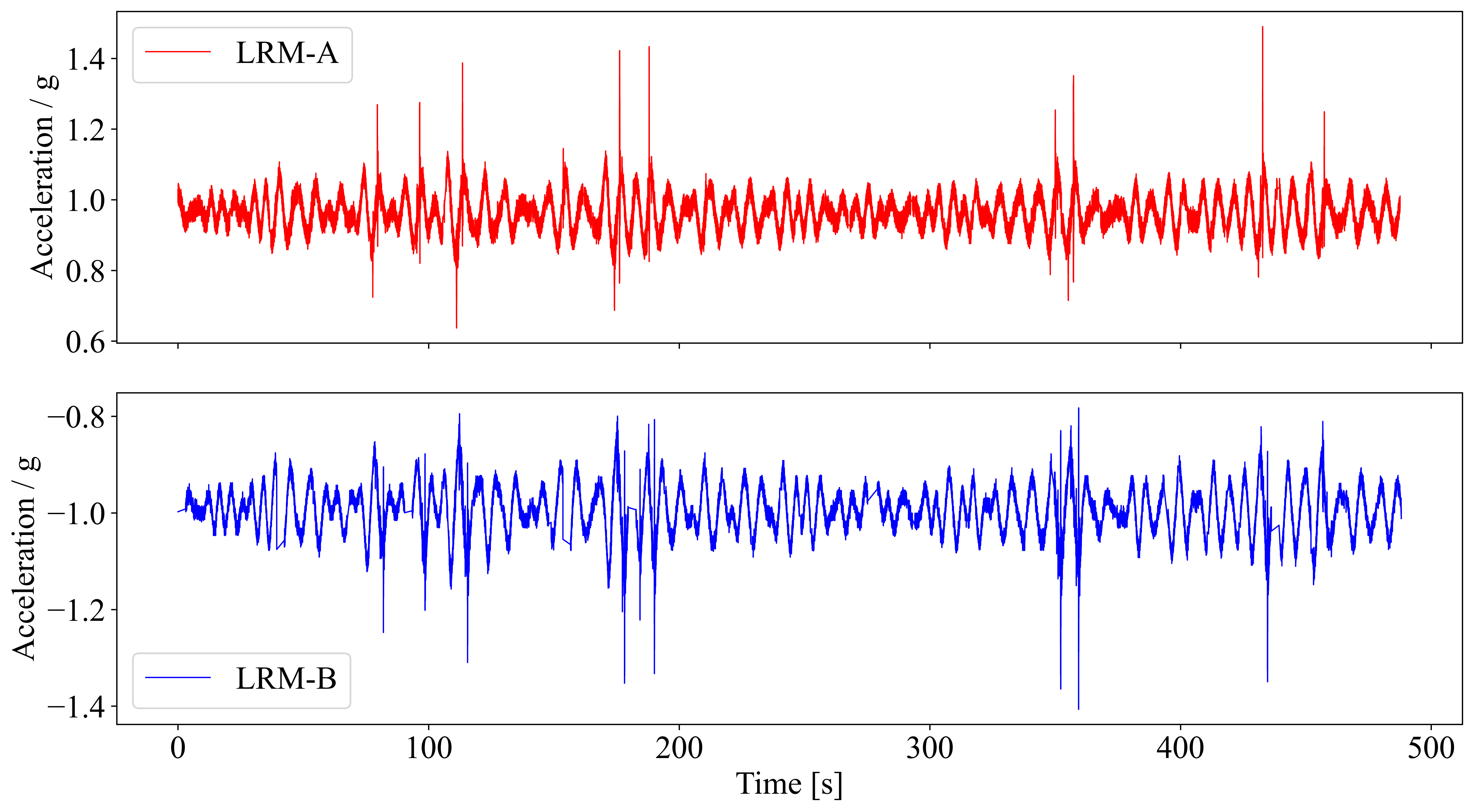}
\caption{\label{fig_sensor_acc} The plot displays sensor data of acceleration in the vertical direction, offering additional evidence of the real-time dynamic stability of the apparatus.}
\end{figure}

In addition, the acceleration in the vertical direction is influenced by the force on the cable and the fluctuation of the research vessel on the sea surface. We observed that the vertical acceleration exhibits periodic behavior, with Fourier analysis revealing a period of about 5 to 10 seconds, as shown in Figure~\ref{fig_sensor_acc}.

\newpage

\section{Refined Model for photon propagation in scatter-less water}
\label{App B}

Building upon the discussion in Section \ref{models}, where we discussed light propagation models, we now aim to establish a refined model for the attenuation effect in a scatter-less water medium. This model considers a point-like or spherically isotropic light source emitting photons uniformly. The light source is positioned at the center of a spherical receiving surface with a radius of $R$. Before the photons are received by the detector, they undergo multiple scattering, leading to an optical path length $L$ that is greater than $R$. The basic idea is to calculate each order of the weighted mean optical path length, $\overline{L}$, for photons that scatters once, twice and multiple times. Then we can incorporate the absorption effect as a weighting term based on the optical path length.

\subsection{The First-Order Scattering Coefficient}

\begin{figure}[htbp]
    \centering
    \includegraphics[width=0.49\linewidth]{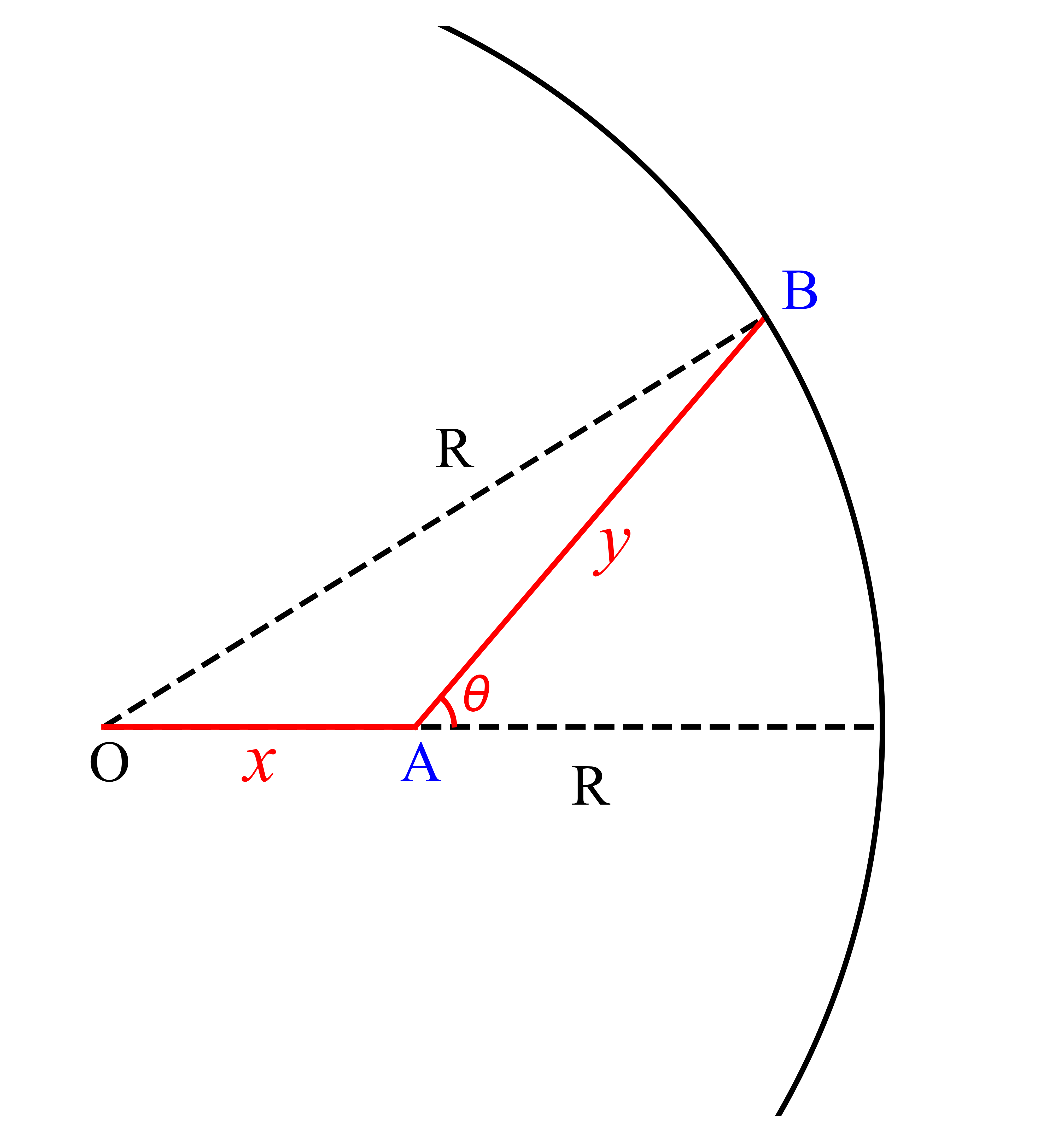}
    \includegraphics[width=0.49\linewidth]{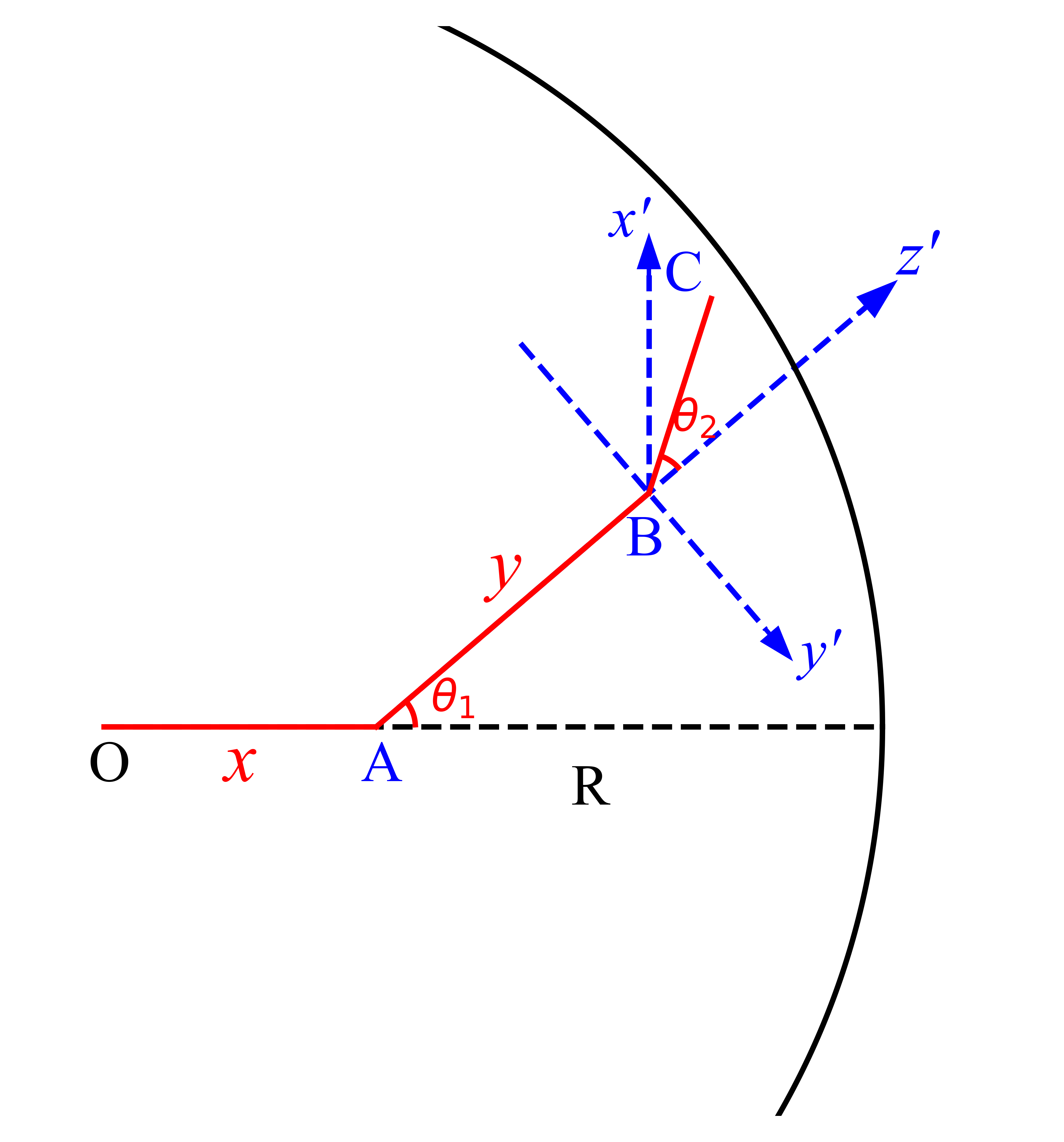}    

    \caption{\label{fig:L1L2} Left panel: optical path of photons that scatter once. Right panel: optical path of photons that scatter twice. } 
\end{figure}

To begin with, we calculate the geometric optical path for photons that scatter once, as depicted in Figure \ref{fig:L1L2}. In this scenario, a photon scatters at a random distance $x$ from the point light source, then propagates a distance $y$ to the detection sphere, at a scattering angle of $\theta$. The mean optical path for these photons can be determined using Formula \ref{L1_bar}, where $L_1 = x+y$. The weight factor $P_{sca}$ is defined by Formula \ref{weight_term}, and $\beta_{sca}(\cos\theta)$ represents the phase function for photons undergoing Mie or Rayleigh scattering. % 加上参考文献17

% L1 光路图

% L1_bar 公式
\begin{equation}
    \overline{L_1}=
    \frac{
        \int_{0}^{R}dx\int_{0}^{\pi}(-\sin\theta)d\theta
    \;L_1(x,\theta) \cdot P_{sca}}{
        \int_{0}^{R}dx\int_{0}^{\pi}(-\sin\theta)d\theta
    \;P_{sca}}
    \label{L1_bar}
\end{equation}

% P_sca 公式
\begin{equation}
    P_{sca} = \beta_{sca}(\cos\theta) \cdot e^{-\frac{L_1(x,\theta)}{\lambda_{sca}}}
    \label{weight_term}
\end{equation}

% Rayleigh and Mie scatter
\begin{equation}
    \beta_{sca}(\cos\theta)
    =\left\{
        \begin{array}{ll}
            \beta_{mie}(\cos\theta)=\frac{1}{2}\frac{1-\mu^2}{(1+\mu^2-2\mu \cos\theta)^\frac{3}{2}}{\quad}{\quad}\mu=\left \langle \mathrm{cos}\theta \right \rangle \\
            \\
            \beta_{ray}(\cos\theta)=\frac{3}{8}(1+\cos^2\theta)
        \end{array}\right. \label{beta}
\end{equation}

Considering the mixture of two kinds of scattering, the weight factor can comprise two components:

\begin{equation}
    P_{mix} = [\frac{\beta_{mie}(\cos\theta)}{\lambda_{mie}}+\frac{\beta_{ray}(\cos\theta)}{\lambda_{ray}}] \cdot e^{-(\frac{1}{\lambda_{mie}}+\frac{1}{\lambda_{ray}})L_1(x,\theta)}
\end{equation}

To simplify the description of this scattering effect, we introduce the definition of the first-order scattering coefficient, $c_{1}$, and have $\overline{L_1}=c_1 R$. We compared this analytically calculated $c_{1}$ with Geant4 simulation results, as shown in Figure~\ref{fig:c1c2}.

It can be seen that the analytical $c_1$ is perfectly matched with the numerical results from Geant4 simulation which verifies the correctness of the first-order scattering average optical path $\overline{L_1}$.

\subsection{The Second-Order Scattering Coefficient}

Similarly, for all the photons scatters twice, the mean optical path can be calculated by building a model shown in Figure \ref{fig:L1L2}, where the photons firstly scatter at the distance of $x$ and at a scattering angle of $\theta_1$, then secondly scatter at another distance of $y$ with a scattering angle of $\theta_2$ and the azimuth angle of $\varphi$. The optical path of the photon can be obtained by:

% \begin{figure}[htbp]
%     \centering
%     \includegraphics[width=0.4\textwidth]{camera_figures/L2.png}
%     \caption{\label{fig:L_2} The geometric diagram of where a photon scatters twice to reach the sphere at the distance of $R$ from the point light source.}
% \end{figure}

\begin{equation}
L_2(x,y,\theta_1,\theta_2,\varphi)=x+y+a(x,y,\theta_1,\theta_2,\varphi)+b(x,y,\theta_1,\theta_2,\varphi)
\end{equation}
where,
\begin{equation}
    a(x,y,\theta_1,\theta_2,\varphi) = -[x(\sin\theta_1\sin\theta_2\sin\varphi+\cos\theta_1\cos\theta_2)+y\cos\theta_2]
\end{equation}
\begin{equation}
    \begin{split}
        b^2(x,y,\theta_1,\theta_2,\varphi) & =  x^2(\sin^2\theta_1\sin^2\theta_2\sin^2\varphi+\cos^2\theta_1\cos^2\theta_2+2\cos\theta_1\cos\theta_2\sin\theta_1\sin\theta_2\sin\varphi-1) \\
        & +xy(2\cos\theta_1\cos^2\theta_2+2\cos\theta_2\sin\theta_2\sin\theta_1\sin\varphi-2\cos\theta_1)+y^2(\cos^2\theta_2-1) \\
        & +R^2
    \end{split}
\end{equation}

% where $x$ is the geometric distance of the photon from the point light source to where the first scattering occurs, $y$ is the geometric distance from where the first scattering to the second scattering, $\theta_1$ and $\theta_2$ indicate the scattering angle of the photon when the first scattering and the second scattering occur, respectively, $\varphi$ is the azimuth angle of the photon when the second scattering occurs. Then the average optical path of all photons that scatter twice is:

\begin{equation}
    \overline{L_2}=
    \frac{
        \int_{0}^{R}dx\int_{0}^{\pi}(-\sin\theta_1)d\theta_1\int_{0}^{f(x,\theta_1)}dy\int_{0}^{\pi}(-\sin\theta_2)d\theta_2\int_{0}^{2\pi}d\varphi
    \;L_2(x,y,\theta_1,\theta_2,\varphi) \cdot P_{sca}}{
        \int_{0}^{R}dx\int_{0}^{\pi}(-\sin\theta_1)d\theta_1\int_{0}^{f(x,\theta_1)}dy\int_{0}^{\pi}(-\sin\theta_2)d\theta_2\int_{0}^{2\pi}d\varphi
    \;P_{sca}}
\end{equation}
Here, the integral interval of $y$ is [0, $f(x,\theta_1)$], where $f(x,\theta_1)$ is given by:.
\begin{equation}
    f(x,\theta_1) = -x\cos\theta_1+\sqrt{R^2-x^2\sin^2\theta_1}
\end{equation}
And the weight factor:

\begin{equation}
    P_{sca} = \beta_{sca}(\cos\theta_1) \cdot \beta_{sca}(\cos\theta_2) \cdot e^{-\frac{L_2(x,y,\theta_1,\theta_2,\varphi)}{\lambda_{sca}}}
\end{equation}
where $\beta_{\mathrm{sca}}(\cos\theta_1)$ and $\beta_{\mathrm{sca}}(\cos\theta_2)$ are phase distribution functions for the first scattering and second scattering, respectively.

% Similarly, we need to modify the weight factor:
% \begin{equation}
%     P_{mix} = [\frac{\beta_{mie}(\cos\theta_1)}{\lambda_{mie}}+\frac{\beta_{ray}(\cos\theta_1)}{\lambda_{ray}}] \cdot [\frac{\beta_{mie}(\cos\theta_2)}{\lambda_{mie}}+\frac{\beta_{ray}(\cos\theta_2)}{\lambda_{ray}}] \cdot e^{-(\frac{1}{\lambda_{mie}}+\frac{1}{\lambda_{ray}})L_2(x,y,\theta_1,\theta_2,\varphi)}
% \end{equation}

\begin{figure}[htbp]
    \centering
    \includegraphics[width=0.49\linewidth]{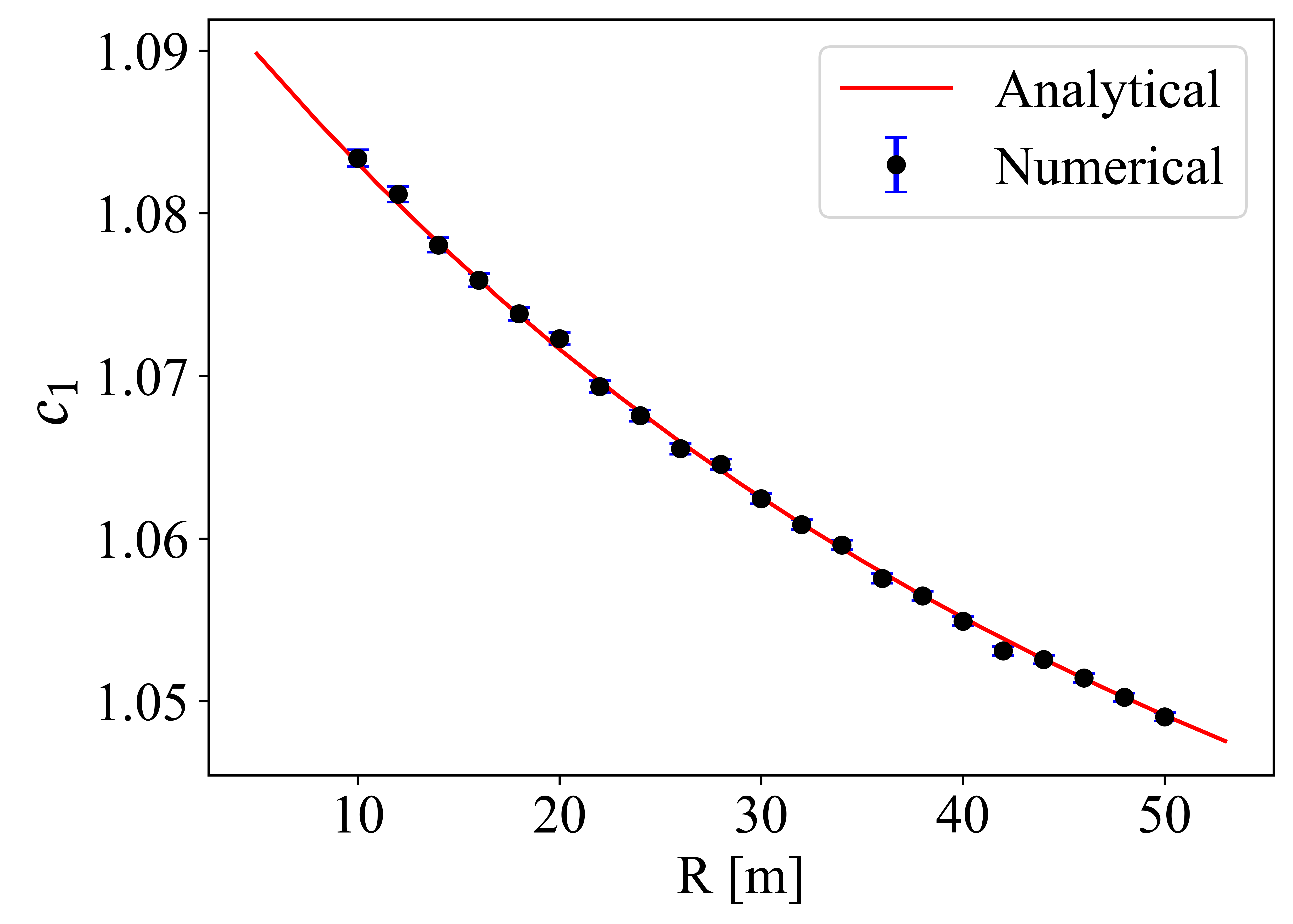}
    \includegraphics[width=0.49\linewidth]{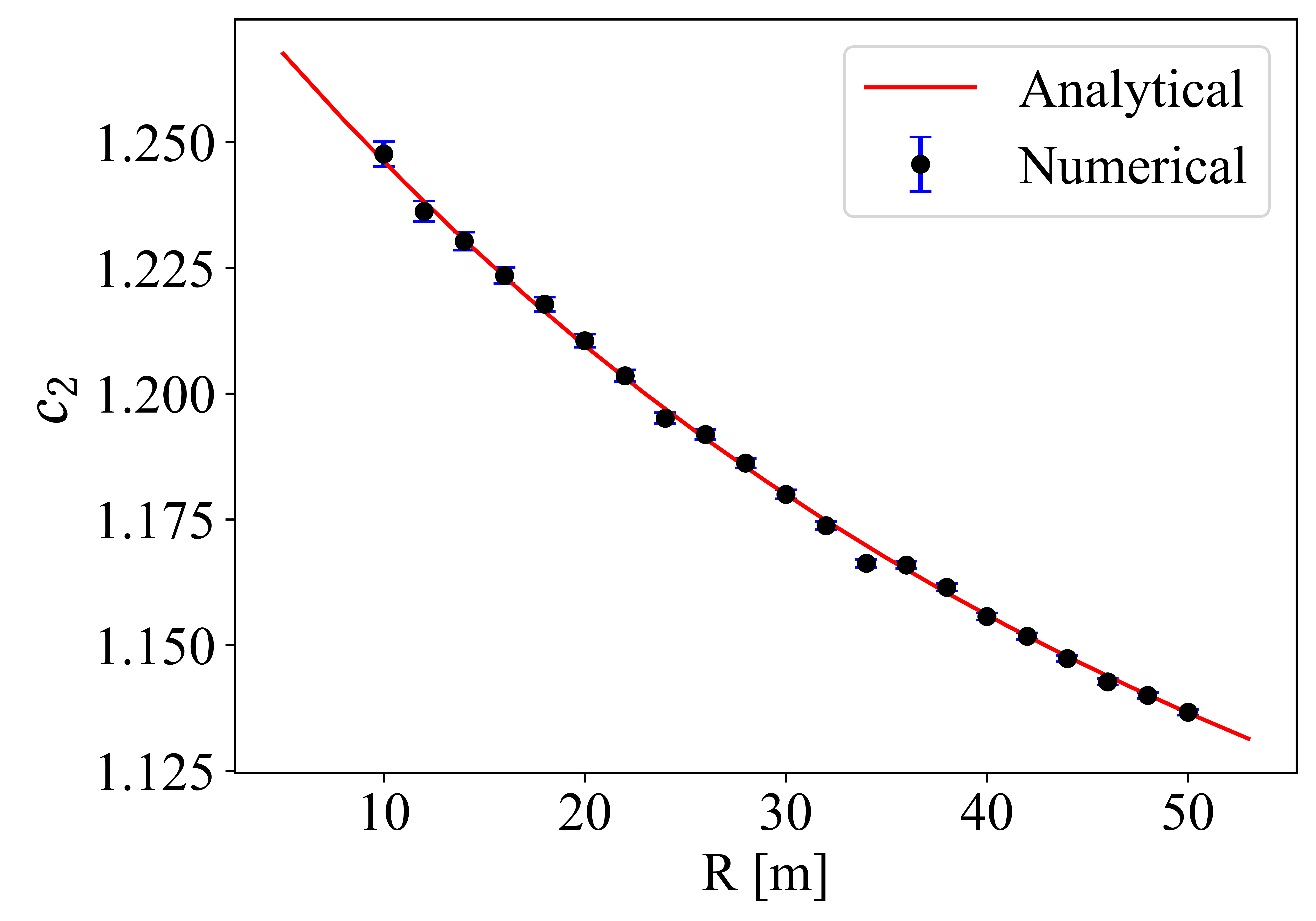}  
    \caption{\label{fig:c1c2} Scatter coefficients with $\lambda_{mie} = 50 m$, $\lambda_{ray} = 200 m$ and $\left \langle \mathrm{cos}\theta \right \rangle = 0.9$. Left panel: $c_1 = \overline{L_1}/L$. Right panel: $c_2 = \overline{L_2}/{L}$.}
\end{figure}

Similarly, the second-order scattering coefficient $c_2$ is defined $\overline{L_2}=c_2 R$. The analytical and numerical solution $c_2$ obtained by analytical solution and Geant4 simulation with the change of $R$ is shown in Figure \ref{fig:c1c2}. 

The conclusion drawn from the model is that each order of scattering effect is associated with a coefficient $c_i$, and the weight factor is influenced by the exponentially decreasing probability of multiple scattering. This model provides valuable insights as it demonstrates that all the essential optical parameters can be derived from the total light intensity detected at different distances. Furthermore, in the context of the PMT system of T-REX, the recorded arrival time of each photon directly corresponds to the average optical path length $\overline{L}$. This enables the absorption length to be easily and independently determined by incorporating a re-weighting factor into the photon counts recorded in each time bin by Cam-A and Cam-B.

% \begin{figure}[htbp]
%     \centering
%     \includegraphics[width=1.0\textwidth]{camera_figures/c2_mix.png}
%     \caption{\label{fig:c_2} In the second-order scattering, both Mie scattering and Rayleigh scattering are possible, and the analytical and numerical solution of $c_2$ change with $R$. The left figure takes the Mie scattering parameter as $\lambda_{mie} = 50 m$ and the Rayleigh scattering parameter as $\lambda_{ray} = 200 m$ with $<\cos> = 0.9$, and the right figure takes the Mie scattering parameter as $\lambda_{mie} = 100 m$ and the Rayleigh scattering parameter as $\lambda_{ray} = 100 m$ with $<\cos> = 0.9$.}
% \end{figure}

\section{The double imaging phenomenon}
\label{App C}

\begin{figure}[htbp]
    \centering
    \includegraphics[width=1\textwidth]{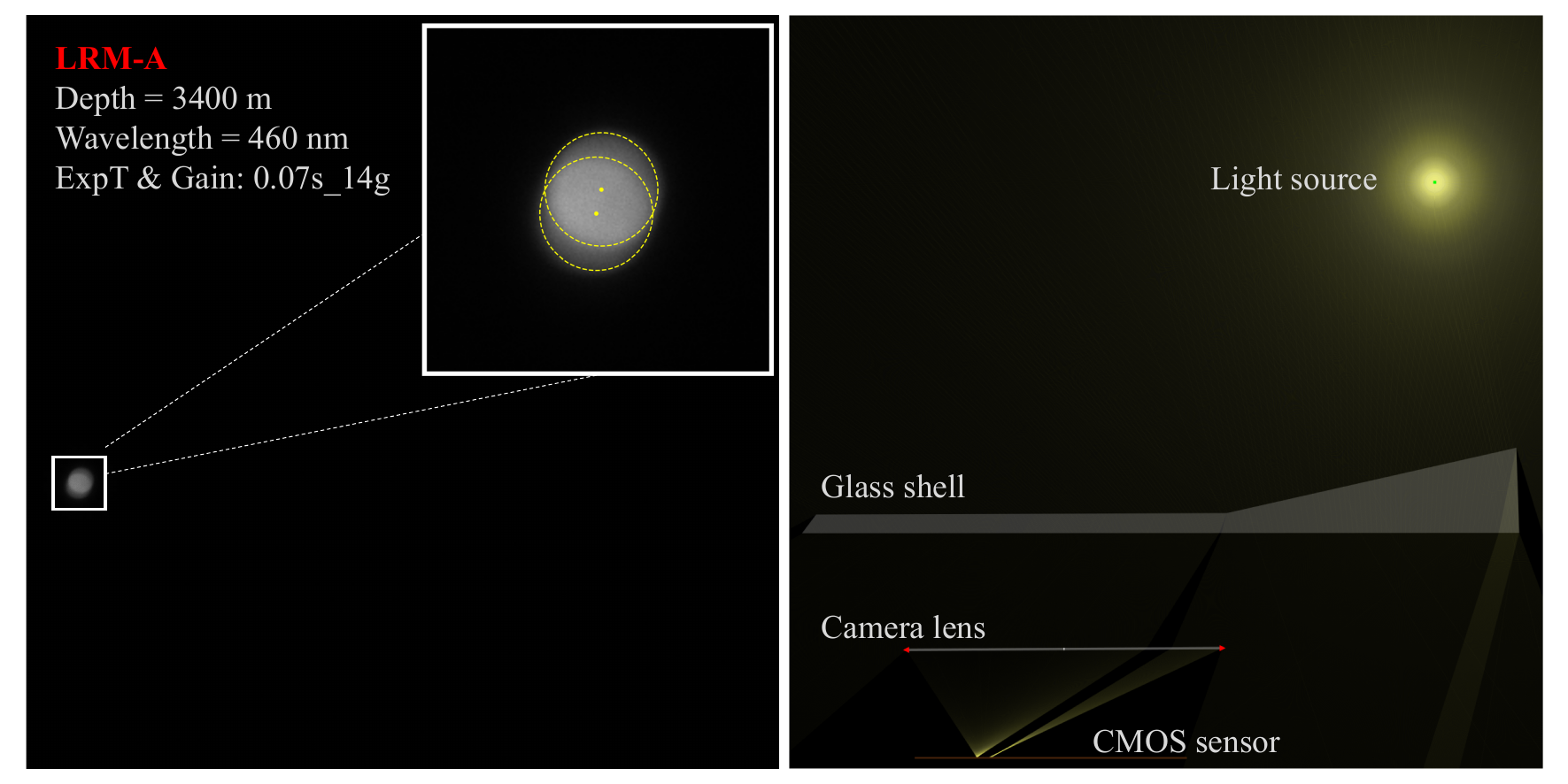}
    % \qquad
    % \includegraphics[width=.4\textwidth,trim=70 530 1350 890,clip]{camera_figures/Ghosting.jpg}
    \caption{\label{fig:Ghosting} The Ghosting Phenomenon of almost two images.}
\end{figure}

During the dynamic T-REX retrieval process, a phenomenon of double imaging was observed in the images captured by LRM-A, as shown in Figure \ref{fig:Ghosting}. This phenomenon occurred when the LEM was nearly out of the viewing field of the LRM-A's camera. At this particular viewing angle, the image appeared as two overlapping circles, with a distance of approximately 15 pixels between their centers. We repeated this phenomenon in subsequent lab tests using LRM-A and estimated the potential uncertainties. The investigation revealed that a small area of the glass surface in front of the camera exhibited unevenness, causing inconsistent refraction of light at certain viewing angles.

% Here, we employed Ray Optics Simulation to illustrate the impact of uneven areas on the light propagation path as we assume that this uneven surface acts as a small-angle slope, as shown in Figure \ref{fig:Ghosting}. By performing geometric optics calculations, we determined that for small slope angles, the difference in imaging distance on the CMOS sensor due to incident light is approximately linear with the slope angle. Utilizing practical parameters, such as the thickness of the glass shell, we found that the slope angle is approximately $1.6^\circ$ when the distance between the centers is 15 pixels, which aligns with our hypothesis. 

To illustrate the impact of uneven areas on the light propagation path, we employed Ray Optics Simulation, assuming that this uneven surface acts as a small-angle slope, as shown in Figure \ref{fig:Ghosting}. Through geometric optics calculations, we determined that for small slope angles, the difference in imaging distance on the CMOS sensor due to incident light is approximately linear with the slope angle. Considering practical parameters such as the thickness of the glass shell, we found that the slope angle is approximately $1.6^\circ$ when the distance between the centers is 15 pixels, confirming our hypothesis.

In conclusion, this double image effect occurs only at a particular angle and contribute negligible uncertainties to the normal image-taking process. However, it provided valuable insights that the application of the camera system requires a smooth optical glass shell.

\newpage
\bibliography{myref}

\end{document}